\documentclass[aps,floatfix,nofootinbib,amsmath,amssymb,showpacs]{revtex4}

\usepackage[dvips]{graphicx}
\usepackage{mathrsfs}

\begin{document} 

\title{Deducing the nature of dark matter from direct and indirect detection 
experiments in the absence of collider signatures of new physics}

\author{Maria Beltran}
\affiliation{Department of Astronomy and 
Astrophysics, The University of Chicago}

\author{Dan Hooper}
\affiliation{Theoretical Astrophysics, Fermi National Accelerator Laboratory}
\affiliation{Department of Astronomy and Astrophysics, 
The University of Chicago}

\author{Edward W.\ Kolb}
\affiliation{Department of Astronomy and Astrophysics, 
Enrico Fermi Institute, and Kavli Institute for Cosmological Physics, 
The University of Chicago}

\author{Zosia A.\ C.\ Krusberg}
\affiliation{Department of Physics, The University of Chicago}

\begin{abstract}

Despite compelling arguments that significant discoveries of physics beyond
the standard model are likely to be made at the Large Hadron Collider, it
remains possible that this machine will make no such discoveries, or will make
no discoveries directly relevant to the dark matter problem. In this article,
we study the ability of astrophysical experiments to deduce the nature of dark
matter in such a scenario. In most dark matter studies, the relic abundance and
detection prospects are evaluated within the context of some specific particle
physics model or models ({\textit e.g.,} supersymmetry).  Here, assuming a 
single weakly interacting massive particle constitutes the universe's dark matter, we attempt to
develop a model-independent approach toward the phenomenology of such 
particles in the absence of any discoveries at the Large
Hadron Collider. In particular, we consider generic fermionic or scalar dark
matter particles with a variety of interaction forms, and calculate the
corresponding constraints from and sensitivity of direct and indirect detection
experiments. The results may provide some guidance in disentangling information
from future direct and indirect detection experiments.

\end{abstract}
\pacs{95.35.+d;95.30.Cq,95.85.Ry,95.55.Ka; FERMILAB-PUB-08-308-A}
\maketitle

\section{Introduction}

The consensus of the astrophysics community is that a large
fraction of the universe's mass consists of non-luminous,
non-baryonic material, known as dark matter
\cite{Bertone:2004pz}. Although the nature of this substance or
substances remains unknown, weakly interacting massive particles
(WIMPs) represent a particularly attractive and well motivated class
of possibilities. Although the most studied WIMP candidate is the
lightest neutralino \cite{neutralino} in supersymmetric models, many
other possibilities have also been proposed, including Kaluza-Klein
states in models with universal \cite{universal,universal2} or warped
\cite{warped} extra dimensions, stable states in Little Higgs theories
\cite{tparity}, and many others.

In each of the above mentioned cases, many new particle species, in
addition to the WIMP itself, are expected to lie within the discovery
reach of the Large Hadron Collider (LHC), making the task of deducing
the nature of the WIMP immeasurably simpler. In supersymmetry, for
example, gluinos and squarks are expected to be produced
prolifically. By studying the cascades produced in the decays of such
particles, the masses of several superparticle masses, including the
lightest neutralino, are likely to be determined. If squarks, gluinos,
and other additional superpartners are too heavy to be produced,
however, the lightest neutralino will also be very difficult to study
at the LHC, even if rather light itself. More generally speaking, in
the absence of heavier particles with shared quantum numbers, WIMPs
will not be easily detected or studied at the LHC. Although an
electroweak scale, cold thermal relic particle, if it exists, would
almost certainly be produced at the LHC, identifying and
characterizing the nature of the WIMP simply from missing energy
studies is a daunting, perhaps impossible,
task \cite{Birkedal:2004xn,Feng:2005gj}.

Although the usual list of prospective WIMPs mentioned above contains
some very attractive and well-motivated candidates for dark matter,
there are certainly many possible forms of dark matter that have not
yet been considered. As the first observations of particle dark matter
might well come from direct and/or indirect detection experiments, it
is possible that these results may be misinterpreted as a result of
theoretical bias, anticipating dark matter to have the properties of a
neutralino or other often studied candidates. To avoid such confusion,
model-independent studies of dark matter phenomenology can play an
important role (for previous work in this direction, see Refs.\
\cite{Birkedal:2004xn,Kurylov:2003ra,Giuliani:2004uk}).

In this article, rather than consider a WIMP candidate from a specific
theoretical model, we study model-independent WIMPs with different
combinations of spins and interaction forms with standard model
particles.  These interactions are limited only by the requirements of
Lorentz invariance and a consequent WIMP abundance consistent with
cosmological observations.  For each spin and interaction form, we
evaluate the constraints from and prospects for direct and
indirect detection of WIMPs in current and future
experiments. Although we will be forced to adopt some assumptions in
order to make the problems at hand tractable, we attempt to be as
general as possible throughout our study. Beyond the starting point
that the dark matter is a WIMP in the form of a single species of a
cold thermal relic, we adopt only two assumptions:

\begin{enumerate} 

\item Any new particle species in addition to the
WIMP has a mass much larger than the WIMP.

\item The WIMP interactions with standard model
particles are dominated by those of one form (scalar, vector,
\textit{etc.}).

\end{enumerate}

An implication of the first assumption is that the WIMP's thermal
abundance is not affected by resonances or coannihilations. At a later
stage of this paper, we will discuss the impact of relaxing these
assumptions.

The remainder of this paper is structured as follows. In Sec.\
\ref{fermion} we explore the phenomenology of a generic fermionic
WIMP, including its annihilation cross section and
relic abundance, elastic scattering cross section and direct detection
prospects, and indirect detection prospects in the form of a neutrino flux
from the Sun and gamma rays and charged particles produced
in galactic annihilations. In Sec.\ \ref{scalar}, we repeat this
exercise for the case of a scalar WIMP. In each of these two sections,
we also consider dark matter candidates from specific particle physics
frameworks and discuss how they fit into our model-independent
analysis. In Sec.\ \ref{conclusion} we summarize our results and
present our conclusions.

\section{Fermionic Dark Matter}\label{fermion}

We begin with the case of a fermionic WIMP, and study five types of
interactions consistent with the requirement of Lorentz invariance. As
mentioned in the Introduction, we assume that the WIMP is the
only new particle at the electroweak scale.  This enables us to
describe the interaction between WIMPs and standard model fermions in
terms of an effective field theory, in which we keep only the first term in the $(q/M)^2$
expansion of the heavy propagator term (here, $q$ and $M$ are the momentum and mass
of the propagator, respectively).  We note that the effective interaction Lagrangians 
are not invariant under the standard model $SU(2)_W \times U(1)_Y$ gauge 
symmetry; however, this is acceptable as our theory need only be valid at energy
scales below the scale of electroweak symmetry breaking.

To begin, we only consider WIMP annihilations to fermion-antifermion pairs,
neglecting for the moment the possibility of final states that include gauge or
Higgs bosons. In particular, we consider the following interaction Lagrangians
between two fermionic WIMPs ($\chi$) and two standard model fermions ($f$):
\begin{eqnarray}
\label{lint-fermionic}
\textrm{Scalar (S):} & \mathscr L & = \frac{G_S}{\sqrt{2}}
\bar \chi \chi \bar f f \\
\textrm{Pseudoscalar (P):} & \mathscr L & = \frac{G_{P}}{\sqrt{2}} 
\bar \chi \gamma^5 \chi \bar f \gamma_5 f \\
\textrm{Vector (V):} & \mathscr L & =\frac{G_V}{\sqrt{2}}
\bar \chi \gamma^{\mu} \chi \bar f \gamma_{\mu} f \\
\textrm{Axial Vector (A):} & \mathscr L & = \frac{G_{A}}{\sqrt{2}}
\bar \chi \gamma^{\mu}\gamma^5 \chi \bar f \gamma_{\mu} \gamma_5 f\\
\textrm{Tensor (T):} & \mathscr L &= \frac{G_T}{\sqrt{2}}
\bar \chi \sigma^{\mu \nu} \chi \bar f \sigma_{\mu \nu} f.
\label{fermionlag}
\end{eqnarray}

We will now proceed to calculate the annihilation cross section, relic
density, and elastic scattering cross sections for a fermionic WIMP.

\subsection{Fermionic WIMP Annihilation and Relic Density}

In each of the cases listed above, we are interested in
determining the cosmological density of WIMPs produced in the early
universe. The first step is to calculate the annihilation cross sections 
to fermion-antifermion pairs as a function of the
Mandelstam variable $s$ for each of the five cases.  The result is
\begin{eqnarray}
    \sigma_S & = & \frac{1}{32\pi} \
    \sum_f G_{S,f}^2 \, c_f \sqrt{\frac{s-4 m_f^2}{s-4 M_\chi^2}} \ 
    \left[\frac{(s-4 M_\chi^2) \ (s - 4 m_f^2)}{s} \right]  \label{sigmaS}\\    
    \sigma_P & = & \frac{1}{32\pi} \
    \sum_f G_{P,f}^2 \, c_f \sqrt{\frac{s-4 m_f^2}{s-4 M_\chi^2}} \ 
    s \\
    \sigma_V  & = & \frac{1}{32\pi} \
    \sum_f G_{V,f}^2 \, c_f \sqrt{\frac{s-4 m_f^2}{s-4 M_\chi^2}} \ 
    \left[s+4M_\chi^2 + \frac{(s-4 M_\chi^2)\ (s-4m^2_f)}{3\, s} + 4m^2_f  
    \right] \\
    \sigma_A  & = & \frac{1}{32\pi} \
    \sum_f G_{A,f}^2 \, c_f \sqrt{\frac{s-4 m_f^2}{s-4 M_\chi^2}} \ 
    \left[s-4M_\chi^2 + \frac{(s-4 M_\chi^2)\ (s-4m^2_f)}{3 \, s} + 4m^2_f 
    \right] \\
    \sigma_T & = & \frac{1}{32\pi} \
    \sum_f G_{T,f}^2 \, c_f \sqrt{\frac{s-4m_f^2}{s-4M_{\chi}^2}} \
    \left[6s + 4M_{\chi}^2 + \frac{8(s-4M_{\chi}^2)(s-4m_f^2)}{3\, s} + 4m_f^2
    \right]
    \label{sigmaAA} 
\end{eqnarray}
where the sum is over the final state fermion species and $c_f$ are
the color factors, equal to 3 for quarks and 1 for leptons. 

To determine the density of relic WIMPs, we solve the Boltzmann equation:
\begin{equation}
\frac{dn_{\chi}}{dt} + 3 H n_{\chi}
= -\langle\sigma |v|\rangle \left[(n_\chi)^2 - (n^\textrm{eq}_\chi)^2 \right], 
\end{equation}
where $H \equiv \dot{a}/a=\sqrt{8\pi\rho/3M_{\rm{Pl}}}$ is the Hubble
rate and $\langle\sigma |v|\rangle$ is the thermally averaged WIMP
annihilation cross section \cite{Gondolo:1990dk}.

In thermal equilibrium, the number density of WIMPs is given by:
\begin{equation}
n^{\rm{eq}}_{\chi} = g \left(\frac{M_{\chi}T}{2\pi}\right)^{3/2} \,
\rm{exp}\left(-\frac{\it M_{\chi}}{\it T}\right),
\end{equation}
where $g=2$ is the number of degrees of freedom of a fermionic WIMP.
At $T \gg M_{\chi}$, the number density of WIMPs was very close to its
equilibrium value and nearly as abundant as any other particle. As the
temperature dropped below $M_{\chi}$, however, the number density was
exponentially suppressed until, eventually, the annihilation and
production rates became much smaller than the expansion rate, and the
species froze out of equilibrium.  Since we are considering cold thermal
relics, freeze-out occurred when WIMPs were non-relativistic and
had velocities much smaller than unity.  Substituting $s \approx 4
M^2_{\chi}+M^2_{\chi} v^2$ to Eqs.\ (\ref{sigmaS}-\ref{sigmaAA}),
and expanding in powers of the relative velocity between two
annihilating WIMPs up to order $v^2$, we find
\begin{eqnarray}
    \sigma_S \vert v \vert  & \approx & \frac{1}{4\pi} \sum_f
    G_{S,f}^2  \, c_f \ M_\chi^2 \ \sqrt{1-m_f^2/M_\chi^2}  \ 
    \left[\frac{1}{4} \left( 1 - \frac{m_f^2}{M_\chi^2}\right) v^2 \right]  
    \label{svs} \\
    \sigma_P \vert v \vert  & \approx & \frac{1}{4\pi} \sum_f
    G_{P,f}^2  \, c_f \ M_\chi^2 \ \sqrt{1-m_f^2/M_\chi^2}  \  
     \\
    \sigma_V \vert v \vert  & \approx & \frac{1}{4\pi} \sum_f
    G_{V,f}^2  \, c_f \ M_\chi^2 \ \sqrt{1-m_f^2/M_\chi^2}  \  
    \left[ \left( 2 + \frac{m_f^2}{M_\chi^2}\right)
    + \frac{1}{12}\left( 1 - \frac{m_f^2}{M_\chi^2}\right) v^2 \right] \\
    \sigma_A \vert v \vert  & \approx & \frac{1}{4\pi} \sum_f
    G_{A,f}^2  \, c_f \ M_\chi^2 \ \sqrt{1-m_f^2/M_\chi^2}  \  
    \left[ \frac{m_f^2}{M_\chi^2} + \frac{1}{12} \left(2-\frac{m_f^2}{M_\chi^2} \right)v^2 \right] \\
    \sigma_T \vert v \vert  & \approx &\frac{1}{4\pi} \sum_f
    G_{T,f}^2 \ c_f \ M_{\chi}^2 \ \sqrt{1-m_f^2/M_\chi^2}  \ 
    \left[ \left( 7 + \frac{m_f^2}{M_\chi^2}\right)
    + \frac{2}{3}\left( 1 - \frac{m_f^2}{M_\chi^2}\right) v^2 \right]
    \label{svt} . 
\end{eqnarray}

Numerical solutions of the Boltzmann equation yield a relic density of 
\cite{KandT}: 
\begin{equation}
\Omega_{\chi} h^2 \approx 
\frac{1.04 \times 10^9 x_F}{M_\textrm{Pl} \sqrt{g_*} (a+3b/x_F)},
\label{sol}
\end{equation}
where $x_F = m_{\chi}/T_F$, $T_F$ is the temperature at freeze-out,
$g_*$ is the number of relativistic degrees of freedom available at
freeze-out ($g_* \approx 92$ for a freeze-out temperature between the
bottom quark and $W$ boson masses), and $a$ and $b$ are terms in the
partial wave expansion of the WIMP annihilation cross section, $\sigma
|v| = a + bv^2 + \mathcal{O}(v^4)$. Evaluation of $x_F$ leads to
\begin{equation}
x_F = \ln\left[c(c+2) \sqrt{\frac{45}{8}}
\frac{g \, M_{\chi} M_{\rm{Pl}}  (a+6b/x_F)}{ 2 \pi^3 \sqrt{g_* (x_F}}\right],
\label{xf}
\end{equation}
where $c$ is an order unity parameter determined numerically. WIMPs
with electroweak-scale masses and couplings generically freeze out at
temperatures in the range of approximately $x_F\approx$ 20 to 30.

\begin{figure}[t]
\centering\leavevmode
\includegraphics[width=3.5in,angle=0]{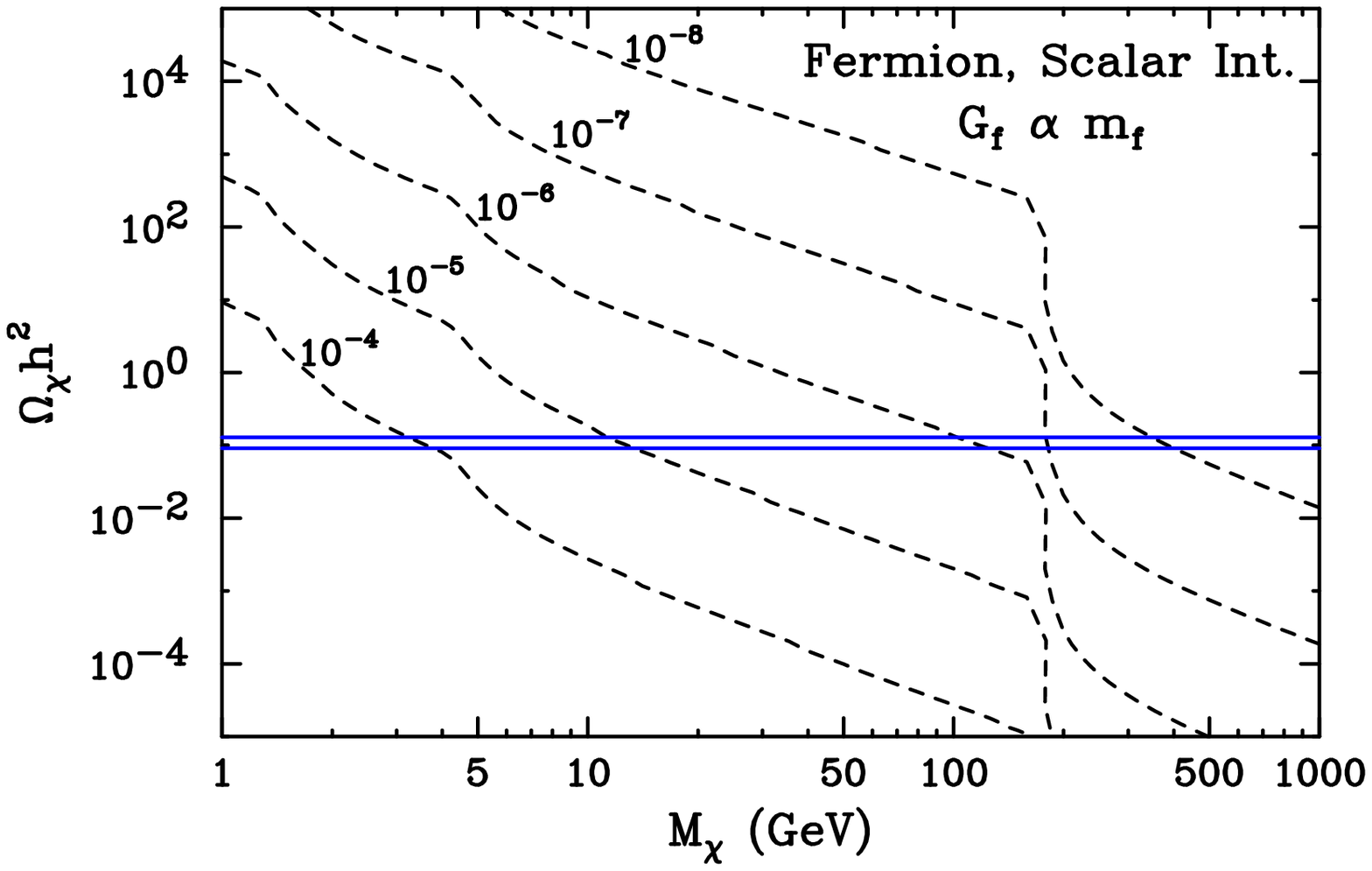}
\includegraphics[width=3.5in,angle=0]{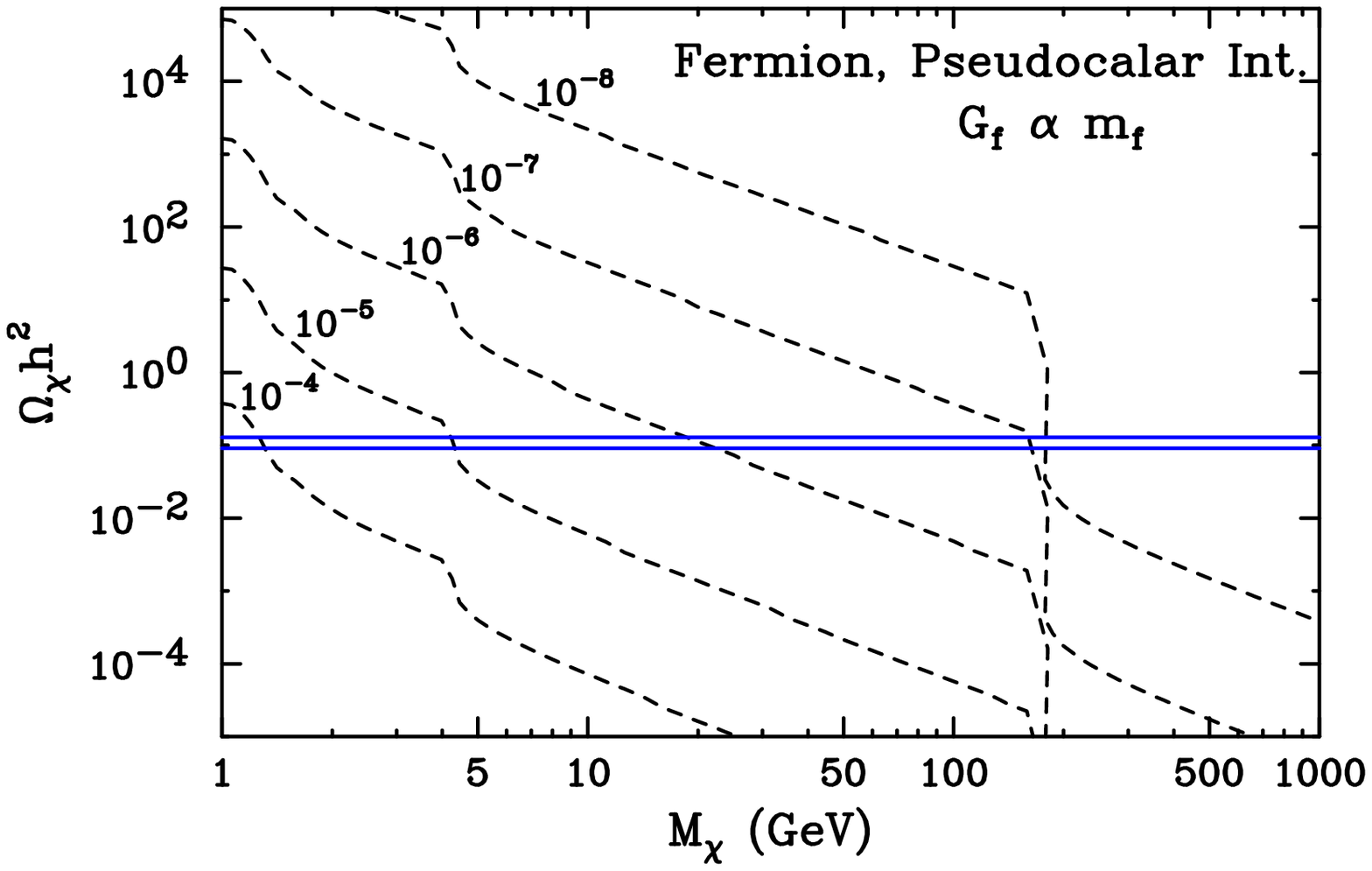}
\includegraphics[width=3.5in,angle=0]{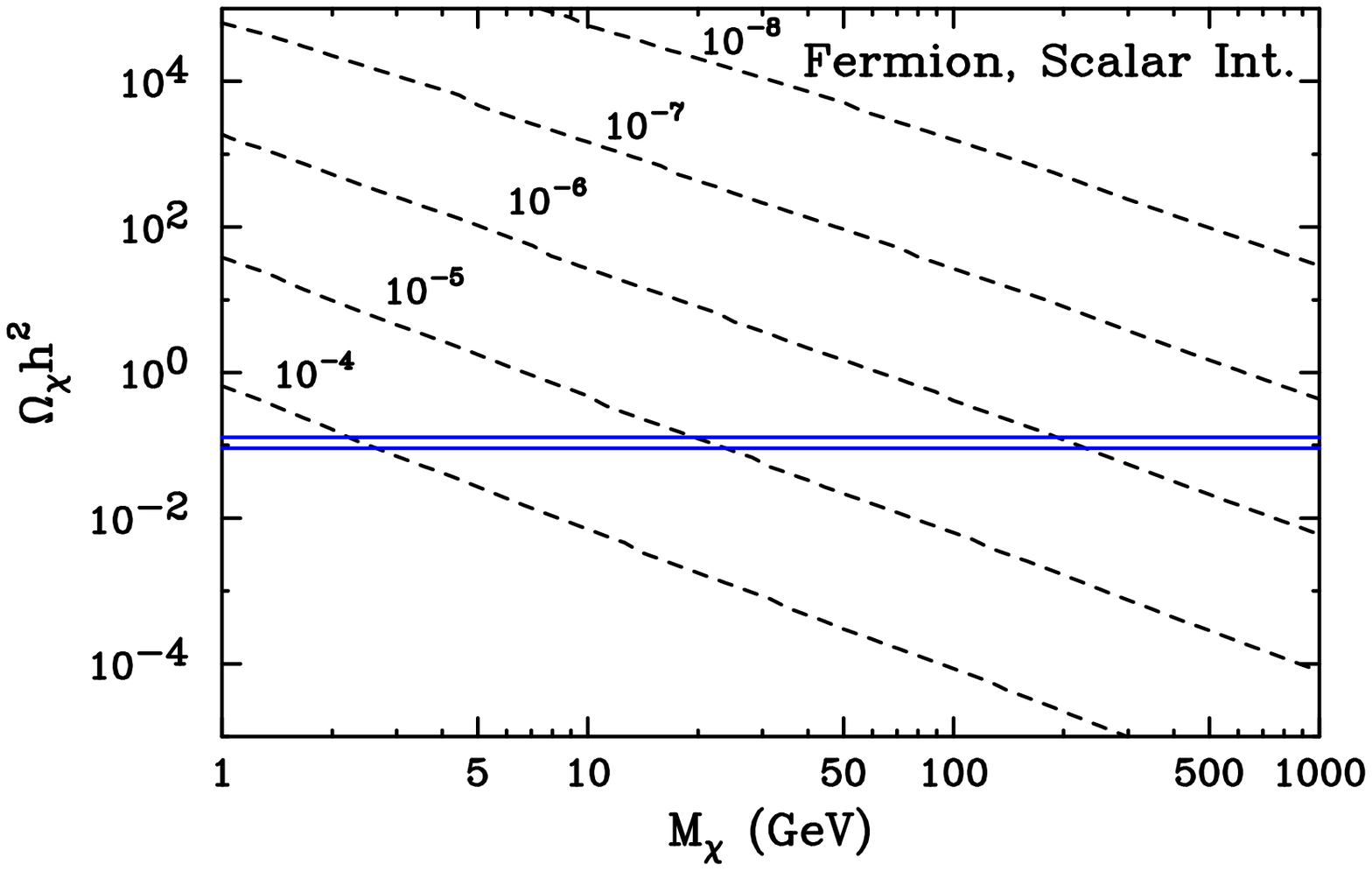}
\includegraphics[width=3.5in,angle=0]{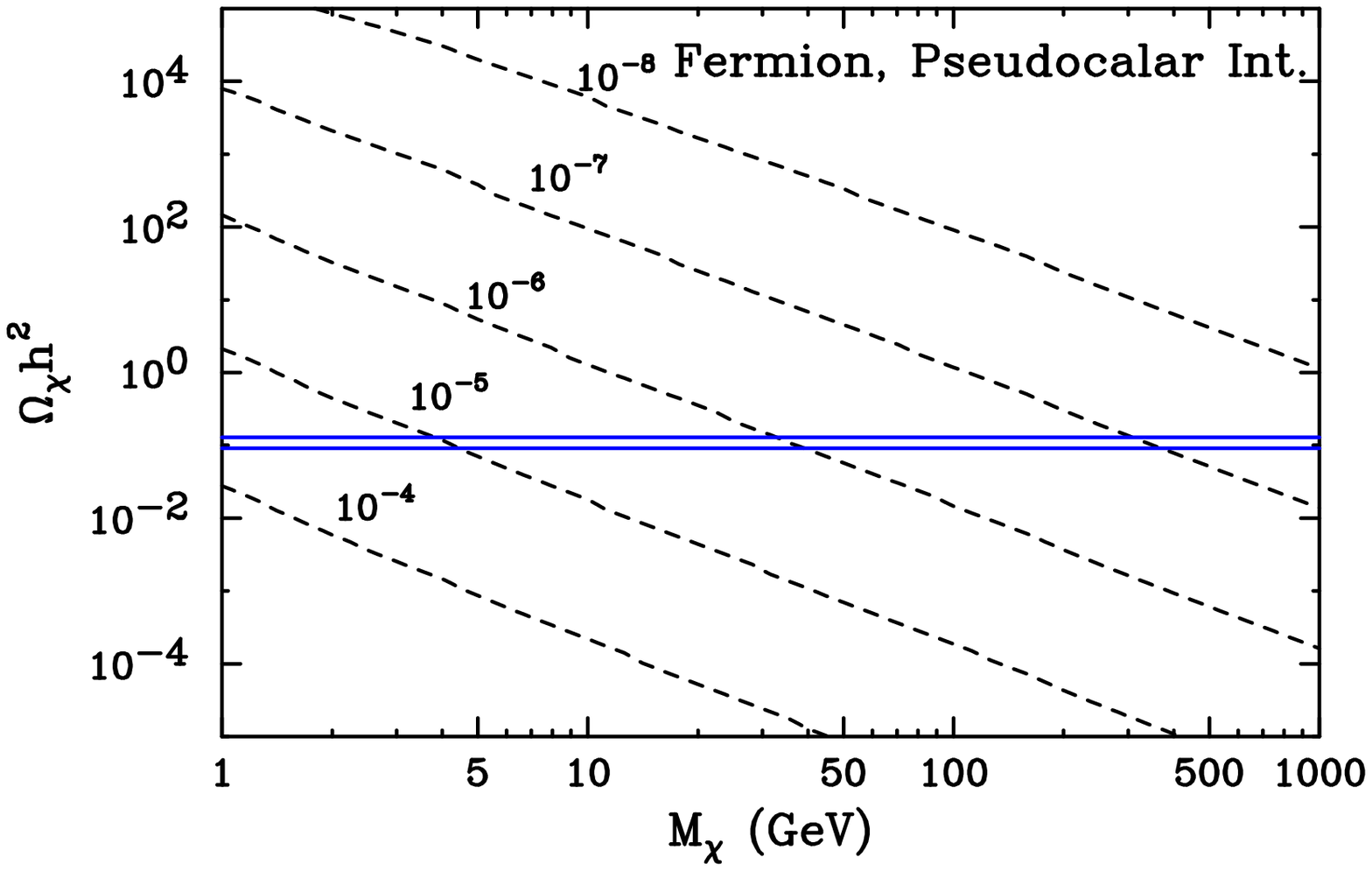}
\includegraphics[width=3.5in,angle=0]{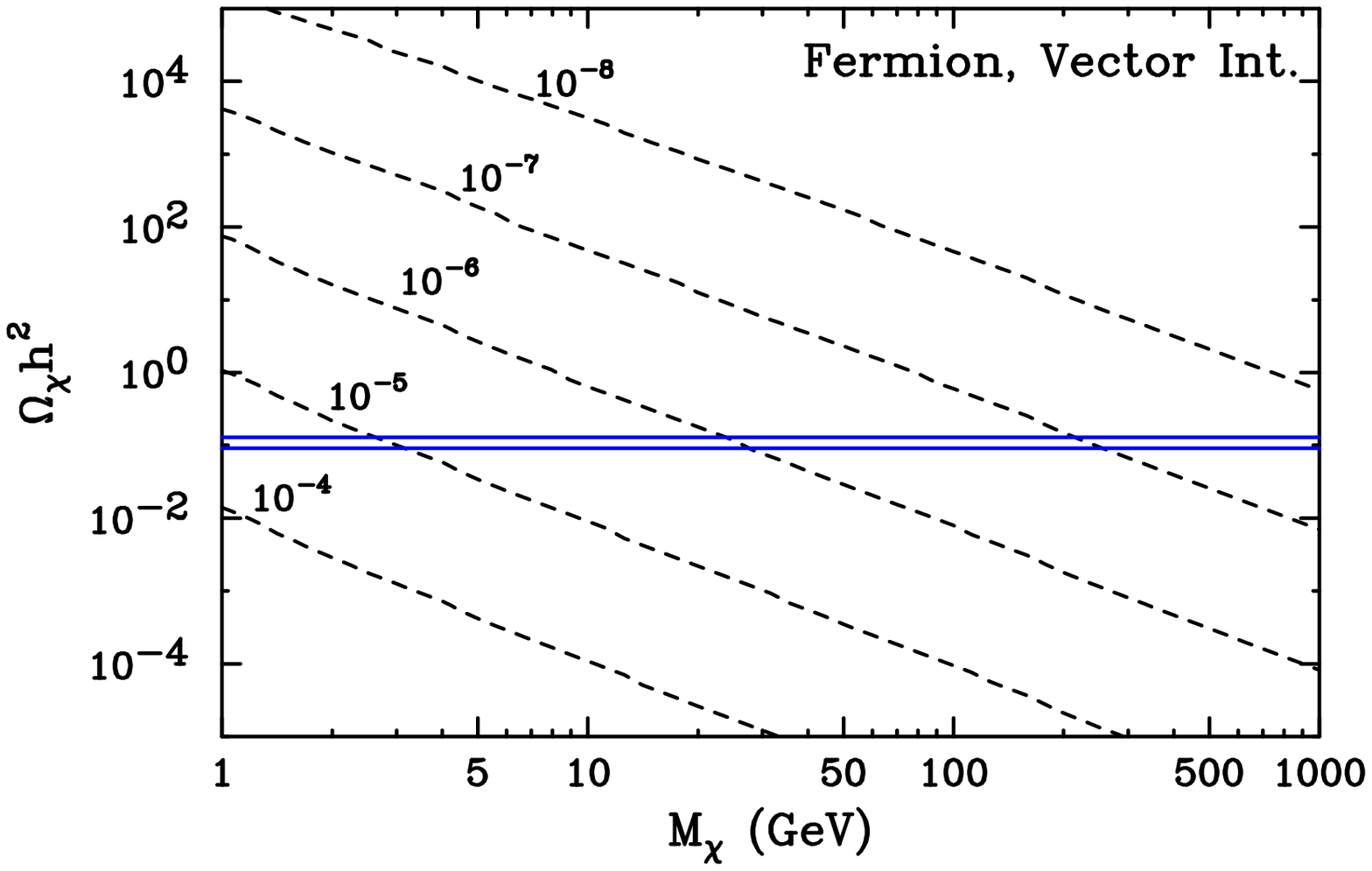}
\includegraphics[width=3.5in,angle=0]{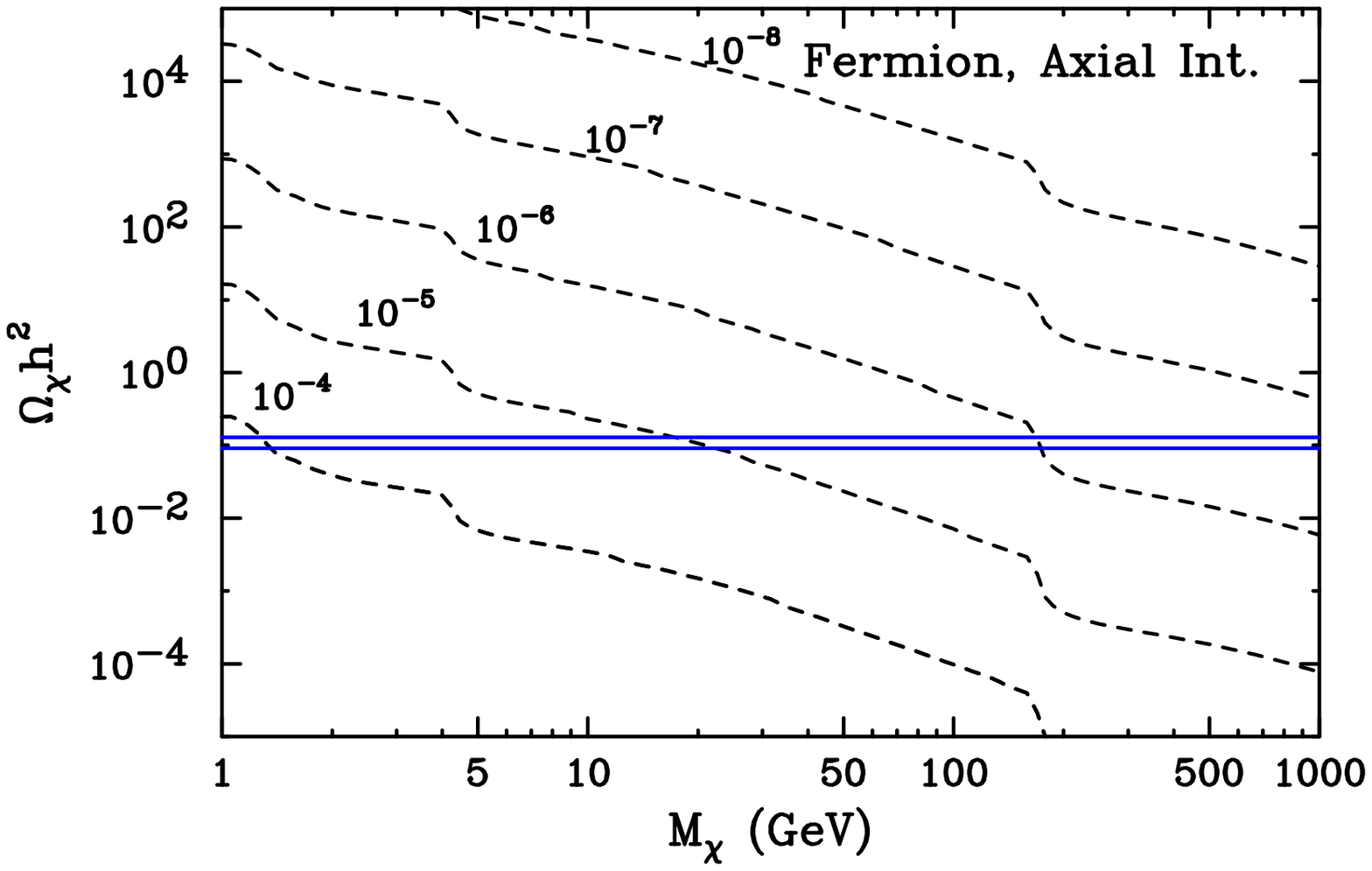}
\caption{The thermal relic density of fermionic dark matter with
scalar, pseudoscalar, vector, and axial interactions. In the upper
left and upper right frames, results are given for effective couplings
to each species of standard model fermion of $G_f \times (1\,{\rm
GeV}/m_f) = 10^{-8}$, $10^{-7}$, $10^{-6}$, $10^{-5}$, and
$10^{-4}$GeV$^{-2}$. In the remaining four frames, results are shown
for $G_f = 10^{-8}$, $10^{-7}$, $10^{-6}$, $10^{-5}$, and $10^{-4}$
GeV$^{-2}$. If resonances, coannihilations, or annihilations to final
states other than fermion-antifermion pairs are significant, the relic
abundance is expected to be significantly lower than shown here. Also
shown as horizontal lines is the range of the cold dark matter density
measured by WMAP \cite{wmap}.}
\label{fermionrelic}
\end{figure}

In the absence of resonances and coannihilations \cite{Griest:1990kh},
an annihilation cross section of $\langle \sigma |v| \rangle \sim 3
\times 10^{-26}$ cm$^3$/s $\approx 1$ pb is required (at the
temperature of freeze-out, $T \sim m_{\chi}/20$) to obtain a relic
abundance in agreement with the dark matter abundance measured by
WMAP, $\Omega_{\chi} h^2 = 0.1099 \pm 0.0062$ \cite{wmap}. Although
the annihilation cross section in the low-velocity limit (relevant to
indirect dark matter searches) is not much lower than this value in
many models, it can be considerably suppressed at low velocities if
terms in the annihilation cross section proportional to $v^2$ dominate
the cross section ({\it i.e.,} if $a \ll b$). Furthermore, if the
depletion of WIMPs in the early universe occurs through resonance
channels or via coannihilations with other states, the low velocity
annihilation cross section can be considerably lower than the value at
freeze-out. For more details regarding the relic density calculation,
see Refs.\ \cite{KandT,Griest:1990kh}.

In Fig.\ \ref{fermionrelic}, we show the thermal relic density of a
fermionic dark matter candidate with scalar, pseudoscalar, vector,
and axial interactions. We do not include a separate figure for the tensor
case, since its annihilation cross section is nearly identical to the vector 
case.  As discussed in the Introduction, these results
were found under the assumptions that a given WIMP's interactions are
dominated by those of only one form (scalar, vector, \textit{etc.}),
that the WIMP's interactions are mediated by particles much heavier
than the WIMP mass (thus avoiding the possibility of resonance
effects), and that the WIMP is considerably lighter than any other new
particles (thus making coannihilations unimportant). Also, we include
only annihilations to fermion-antifermion pairs (neglecting the
possibility of final states including gauge or Higgs bosons).

In each frame of Fig.\ \ref{fermionrelic}, we show the relic density
for various values of the effective couplings. In the upper left and
upper right frames, we show results for couplings of $G_f \times
(1\,{\rm GeV}/m_f) = 10^{-8}, 10^{-7}, 10^{-6}, 10^{-5},\, {\rm and}
\, 10^{-4}$ GeV$^{-2}$. This proportionality of the couplings to the
fermion mass is predicted for Yukawa couplings of a Higgs mediated
interaction, for example. In the remaining four frames, we show
results for the case of universal couplings, $G_f = 10^{-8}, 10^{-7},
10^{-6}, 10^{-5},\, {\rm and} \, 10^{-4}$ GeV$^{-2}$.

If any of our assumptions are broken, the resulting thermal relic
abundance will be altered as well. In particular, resonances (or more
generally, a departure from $2 M_{\chi} \ll M_{\psi}$) or
coannihilations could potentially reduce the abundances shown in Fig.\
\ref{fermionrelic} considerably. Additionally, annihilations to final
states such as gauge or Higgs bosons, if significant, could also
reduce the relic density. The effective couplings described in
Fig.\ \ref{fermionrelic} that lead to the correct relic abundance,
therefore, can be thought of as approximate maximal values
allowed for a thermal WIMP.\footnote{Larger couplings may be possible if the density of WIMPs is enhanced by post-freezeout decays of other particles or other non-thermal production mechanisms. For example, see \cite{universal2}.}  Smaller couplings are possible if
appropriate departures are made from our set of assumptions.

\subsection{Direct Detection}
\label{directfermion}

Although only weakly coupled to baryons, WIMPs can occasionally
scatter elastically with atomic nuclei, providing the potential for
detection. Direct detection experiments attempt to measure the recoil
energies of nuclei resulting from such interactions. The interactions
leading to the elastic scattering of WIMPs with nuclei can be
classified as either spin-independent or spin-dependent. In the former
case, WIMPs scatter coherently with an entire nucleus, leading to a
cross section that scales with the square of the atomic number of the
target nuclei. In the later case, the WIMP couples to the spin of the
target nucleus. In the relevant non-relativistic limit, scalar, vector, and
tensor couplings result in a spin-independent interaction, whereas
axial couplings lead to a spin-dependent
interaction \cite{reptJungman}. In this subsection, we focus on the
spin-independent elastic scattering of WIMPs with nuclei, as the
direct detection constraints for this class of interactions are
considerably more stringent. In the next subsection, we will return to
spin-dependent scattering within the context of WIMP capture in the
Sun.

The WIMP-nucleus cross section for spin-independent elastic scattering
is given by
\begin{equation}
\label{scattering-fs}
\sigma_{\chi N}=\frac{4}{\pi}\frac{M_{\chi}^2m_N^2}{(M_{\chi}+m_N)^2}
\left[Z f_p+(A-Z)f_n\right]^2 
\end{equation} 
where $A$ and $Z$ are the atomic
mass and atomic number of the target nuclei.  The effective couplings
to protons and neutrons, $f_{p,n}$, can be written in terms of the
WIMP's couplings to quarks. In the case of a scalar interaction
\begin{equation}
\label{fpn}
f_{p,n}=\sum_{q=u,d,s} \frac{G_{q}}{\sqrt{2}} f^{(p,n)}_{Tq}
\frac{m_{p,n}}{m_q}+\frac{2}{27}f^{(p,n)}_{TG}\sum_{q=c,b,t}
\frac{G_{q}}{\sqrt{2}} \frac{m_{p,n}}{m_q}, 
\end{equation} 
where $G_q$ denotes the WIMP's effective Fermi coupling for a given quark
species. The first term reflects scattering with light quarks, while the second
term accounts for interactions with gluons through a heavy quark loop. The
values of $f^{(p,n)}_{T_q}$ are proportional to the matrix element, $\langle
\bar q q\rangle$, of quarks in a nucleon and have been measured to be
$f^{p}_{Tu}=0.020\pm0.004$, $f^{p}_{Td}=0.026\pm0.005$,
$f^{p}_{Ts}=0.118\pm0.062$, $f^{n}_{Tu}=0.014\pm0.003$,
$f^{n}_{Td}=0.036\pm0.008$, $f^{n}_{Ts}=0.118\pm0.062$ \cite{Ellis:2000ds}. The
value of $f^{(p,n)}_{TG}$ is given by
$f^{(p,n)}_{TG}=1-\sum_{u,d,s}f^{(p,n)}_{Tq}$ and is approximately $0.84$ and
$0.83$ for protons and neutrons, respectively.

In the case of a Yukawa-like scalar interaction ($G_{q} \propto m_q$),
there are significant contributions from both light and heavy
quarks. In the case in which the ratio of the effective scalar
coupling to the quark mass, $G_{S,q}/m_q$, is the same for each quark
species, we arrive at a cross section per nucleon of
\begin{equation}
\sigma_{\chi, p} \approx 3 \times 10^{-7} \, \textrm{pb} \, \times
\left[\frac{G_{S,q} \times (1 \, \textrm{GeV}/m_q)}{10^{-7} \,
\textrm{GeV}^{-2}}\right]^2.  
\end{equation}

In contrast, if we consider the case in which the scalar couplings to
all quarks types are equal (universal couplings), the resulting cross
section is much larger: 
\begin{equation} 
\sigma_{\chi, p} \sim 3 \times 10^{-4} \,
\textrm{pb} \, \times \left(\frac{G_{S,q}}{10^{-7} \,
\textrm{GeV}^{-2}}\right)^2.
\label{universalelastic}
\end{equation}

\begin{figure}[t]
\centering\leavevmode
\includegraphics[width=3.5in,angle=0]{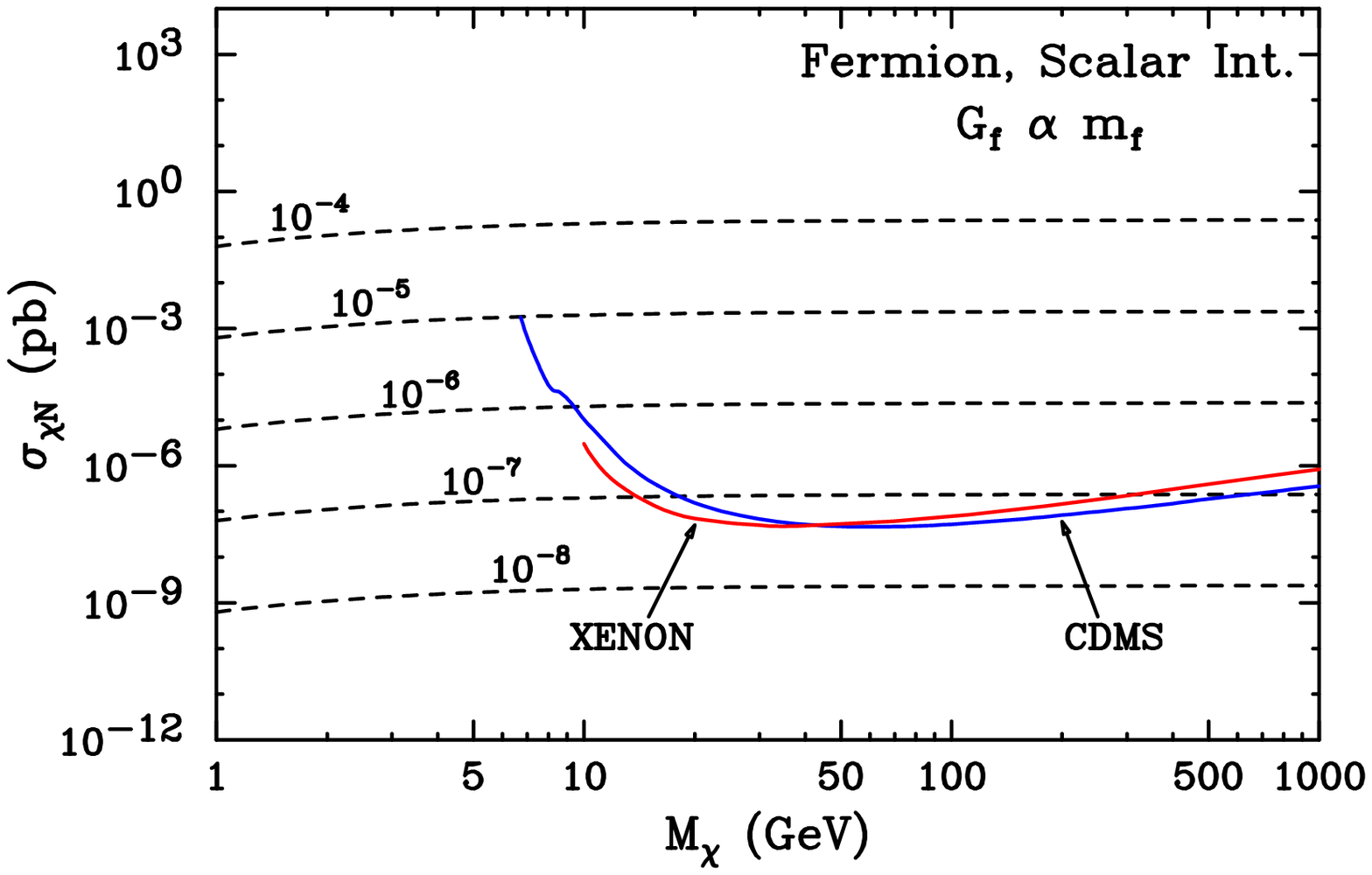}
\includegraphics[width=3.5in,angle=0]{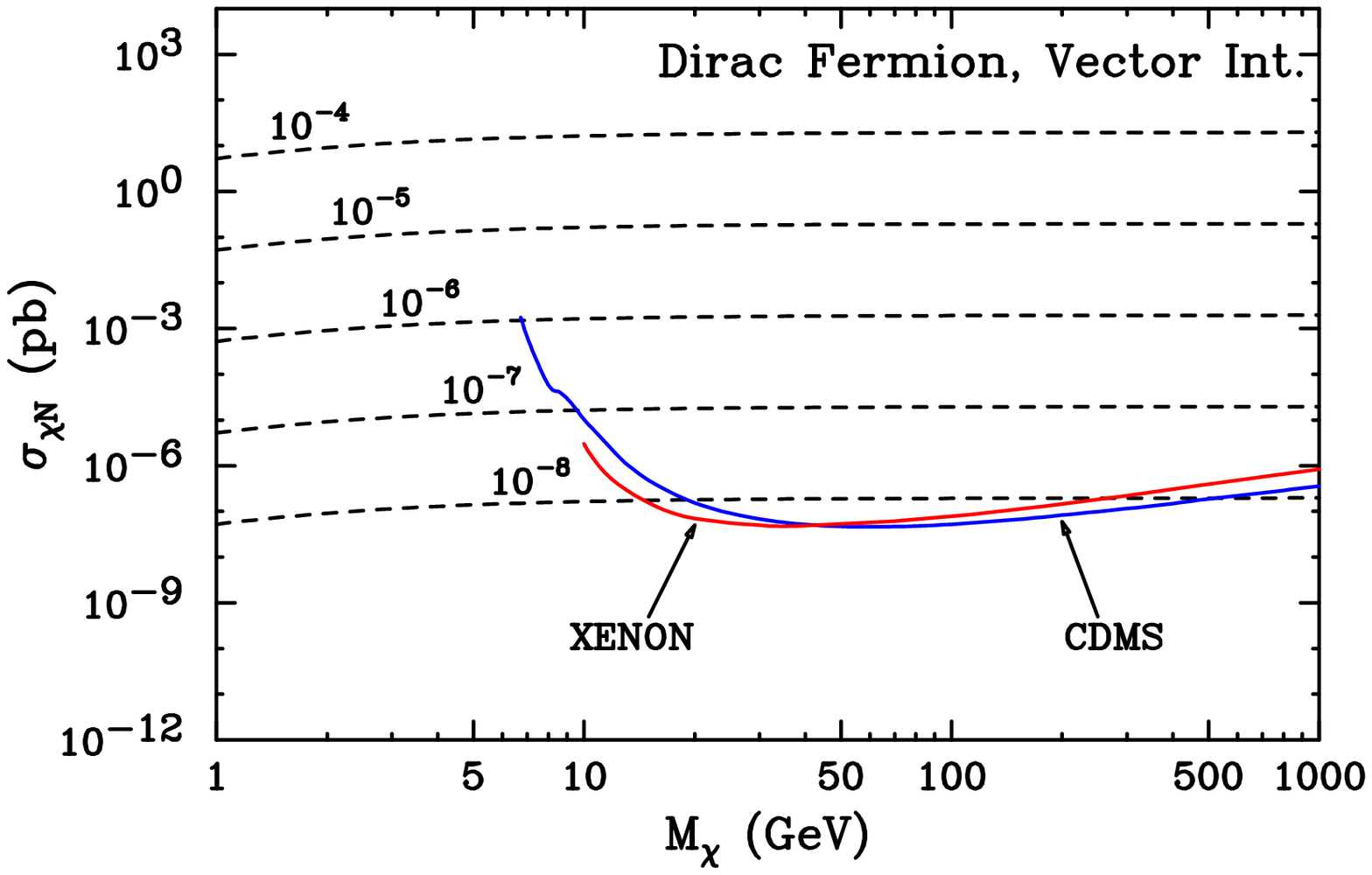}
\caption{The spin-independent WIMP-nucleon elastic scattering cross
section as a function of WIMP mass for a fermionic WIMP interacting
through scalar (left) and vector (right) interactions. Results are
given for effective scalar couplings to each quark species of $G_q
\times (1\,{\rm GeV}/m_q) = 10^{-8}, 10^{-7}, 10^{-6}, 10^{-5}, {\rm
and} \, 10^{-4}$ GeV$^{-2}$ and for effective vector couplings to each
quark of $G_q = 10^{-8}, 10^{-7}, 10^{-6}, 10^{-5}, {\rm and} \,
10^{-4}$ GeV$^{-2}$. Also shown as solid curves are the current upper
limits from the CDMS \cite{cdms} and XENON \cite{xenon} experiments.
We do not show the case in which the scalar couplings are equal for
each quark species, as its leads to much larger cross sections and are
strongly excluded.}
\label{fermionSI}
\end{figure}

The cross section for the scalar case of $G_q \propto m_q$ is shown in
the left frame of Fig.\ \ref{fermionSI}, and compared to the current
upper limits from the CDMS \cite{cdms} and XENON \cite{xenon}
experiments. For fermionic WIMPs heavier than about $10$ GeV, scalar
couplings are constrained to be smaller than $G_{S,q} \times (1\,{\rm
GeV}/m_q) \sim 10^{-7}$. Comparing this result to those shown in Fig.\
\ref{fermionrelic}, we find that fermionic WIMPs must either be
heavier than the top quark threshold to avoid being overproduced in
the early universe {\it and} avoid direct detection constraints {\it
or} some combination of resonance annihilation, coannihilation, or
annihilations to final states other than quarks must dominate the
thermal freeze-out process.

If instead we were to consider the case of universal scalar couplings
to all quark types, as applied in Eq.\ (\ref{universalelastic}), we find
an even more stringent constraint. In particular, the entire range of
couplings that could potentially lead to an acceptable relic density
is excluded by current direct detection constraints by multiple orders
of magnitude. We therefore conclude that if a WIMP is to annihilate
largely through scalar interactions, its couplings to light quarks
must be considerably suppressed (such as in the case of Yukawa-like
couplings, $G_f \propto m_f$) if it is to avoid being excluded by
current direct detection constraints.

In contrast to scalar interactions, pseudoscalar interactions do not
lead to a significant elastic scattering cross section between WIMPs
and nucleons in the low velocity limit. The reason for this can be
seen if one explicitly computes the quark contribution of the vertex,
$\bar q \gamma_5 q$, which goes to zero in the limit of zero momentum
\cite{Halzen:1984mc}. The same conclusion is reached in Ref.\
\cite{Kurylov:2003ra}, in which the relevant nuclear matrix elements
are calculated.

In the case of a Dirac (non-Majorana) fermion, a vector coupling can
also generate a spin-independent elastic scattering cross section. In
contrast to the scalar case, a vector interaction will be dominated by
couplings to the up and down quarks in the nucleon:
\begin{eqnarray}
\label{f-factors}
f_{p}=2 \, \frac{G_{V,u}}{\sqrt{2}} +  \frac{G_{V,d}}{\sqrt{2}}, 
\,\,\,\,\,\, f_{n}=\frac{G_{V,u}}{\sqrt{2}} + 2 \, \frac{G_{V,d}}{\sqrt{2}}.
\end{eqnarray}

If we assume $G_{V,u} \approx G_{V,d}$, this leads to a
spin-independent elastic scattering cross section (per nucleon) of
$\sigma_{\chi, p} \approx 2\times 10^{-5} \, \rm{pb} \times
(G_V/10^{-7} \, \rm{GeV}^{-2})^2$. From the right frame of Fig.\
\ref{fermionSI}, we see that this cross section is in excess of
current experimental limits \cite{cdms,xenon} unless $G_V \lesssim
10^{-8}$ GeV$^{-2}$. Comparing this to Fig.\ \ref{fermionrelic},
however, we find that in order for a Dirac fermionic WIMP with a mass
in the range 10 to 1000 GeV to annihilate largely through a vector
interaction, it must be depleted in the early universe by some
combination of resonance annihilation, coannihilation, or
annihilations to final states other than quarks if it is to avoid
direct detection constraint without being overproduced in the early
universe.  This conclusion also holds for a fermionic WIMP with a
tensor interaction.

We would like to emphasize that the elastic scattering cross sections
we have calculated here should be thought of as approximate upper
limits (again, assuming no late time decays or other non-thermal mechanisms are responsible for the dark matter density). If coannihilations, resonances, or annihilations to leptons,
gauge or Higgs bosons dominated the freeze-out process, then the
effective couplings required to generate the observed relic abundance
may be considerably smaller, leading to reduced elastic scattering
cross sections with nuclei.

\subsection{Neutrinos From WIMP Annihilations In The Sun}

If WIMPs accumulate in the core of the Sun in sufficient numbers,
their annihilations can potentially produce an observable flux of high-energy 
neutrinos \cite{neutrinos}. WIMPs in the Solar System
elastically scatter with nuclei in the Sun and become gravitationally
bound at the rate approximately given by \cite{capture}
\begin{equation}
C_{\odot} \approx 3.35 \times 10^{19} \, \mathrm{s}^{-1} 
\left( \frac{\sigma_{\chi-p,\mathrm{SD}} +\, \sigma_{\chi-p,\mathrm{SI}}
+ 0.07 \, \sigma_{\chi-\mathrm{He,SI}}     } {10^{-7}\, \mathrm{pb}} \right)
\left( \frac{100 \, \mathrm{GeV}}{m_{\chi}} \right)^2 ,
\label{capture}
\end{equation}
where $\sigma_{\chi-p,\mathrm{SD}}$, $\sigma_{\chi-p,\mathrm{SI}}$
and $\sigma_{\chi-\mathrm{He,SI}}$ are the spin-dependent (SD) and
spin-independent (SI) elastic scattering cross sections of WIMPs with
hydrogen (protons) and helium nuclei, respectively. The factor of
$0.07$ reflects the solar abundance of helium relative to hydrogen and
well as dynamical factors and form factor suppression.

The number of WIMP in the Sun, $N$, evolves as
\begin{equation}
\dot{N} = C_{\odot} - A_{\odot} N^2  ,
\end{equation}
where $A_{\odot}$ is the WIMP's annihilation cross section times the
relative velocity divided by the effective volume of the Sun's
core. The present annihilation rate in the Sun is given by
\begin{equation} 
\Gamma = \frac{1}{2} A_{\odot} N^2 = \frac{1}{2} \, C_{\odot} \, 
\tanh^2 \left( \sqrt{C_{\odot} A_{\odot}} \, t_{\odot} \right) \;, 
\end{equation}
where $t_{\odot} \approx 4.5$ billion years is the age of the solar
system.  The annihilation rate is maximized when it reaches
equilibrium with the capture rate ({\it i.e.,} when $\sqrt{C_{\odot}
A_{\odot}} t_{\odot} \gg 1$).  These WIMP annihilations lead to a flux
of neutrinos at Earth given by
\begin{equation}
\frac{dN_{\nu_{\mu}}}{dE_{\nu_{\mu}}} = 
\frac{C_{\odot} F_{\rm{Eq}}}{4 \pi D_{\odot-\oplus}^2}   
\left(\frac{dN_{\nu}}{dE_{\nu}}\right)^{\rm{Inj}},
\label{wimpflux}
\end{equation}
where $C_{\odot}$ is the capture rate of WIMPs in the Sun,
$F_{\rm{Eq}}$ is the non-equilibrium suppression factor (approximately $1$
for capture-annihilation equilibrium), $D_{\odot-\oplus}$ is the Earth-Sun
distance and $(dN_\nu/dE_\nu)^{\rm{Inj}}$ is the neutrino
spectrum from the Sun per WIMP annihilating, which depends on the mass
of the WIMP and its dominant annihilation modes. Due to
$\nu_{\mu}-\nu_{\tau}$ vacuum oscillations, the muon neutrino flux
observed at Earth is the average of the $\nu_{\mu}$ and $\nu_{\tau}$
components.

Muon neutrinos produce muons in charged current interactions with
nuclei in the material inside or near the detector volume of a high-energy 
neutrino telescope. The rate of neutrino-induced muons observed
in a high-energy neutrino telescope is given by
\begin{equation}
N_{\rm{events}} \approx \int \int \frac{dN_{\nu_{\mu}}}{dE_{\nu_{\mu}}}\, 
\frac{d\sigma_{\nu}}{dy}(E_{\nu_{\mu}},y) \,R_{\mu}((1-y)\,E_{\nu})\, 
A_{\rm{eff}} \, dE_{\nu_{\mu}} \, dy,
\end{equation}
where $d\sigma_{\nu}/dy(E_{\nu_{\mu}},y)$ is the neutrino-nucleon
charged current interaction cross section, $(1-y)$ is the fraction of
neutrino energy that goes into the muon and $A_{\rm{eff}}$ is the
effective area of the detector. The factor $R_{\mu}$ is either the distance a
muon of energy $E_{\mu}=(1-y)\,E_{\nu}$ travels before falling below
the muon energy threshold of the experiment, called the muon range, or
the width of the detector, whichever is larger.  The spectrum and flux
of neutrinos generated in WIMP annihilations is determined by the
WIMP's mass and leading annihilation modes.

If the rate at which WIMPs are captured in the Sun is dominated by
spin-independent scattering, one can translate the bounds from CDMS
\cite{cdms} and XENON \cite{xenon} into an upper limit on the neutrino
flux. In fact, even for the maximum elastic scattering cross section
allowed by these experiments, no more than a few neutrino-induced
muons will be generated per year in a kilometer-scale detector
\cite{darkhalzen}. This is well below the sensitivity of next
generation neutrino telescopes such as IceCube \cite{icecube}. Thus, if
we are to detect WIMP annihilations using neutrino telescopes, the
capture rate must be dominated by spin-dependent scattering, which is
far less constrained by direct detection experiments.

The WIMP-nucleus spin-dependent elastic scattering cross section is
approximately given by \cite{reptJungman}
\begin{equation}
\label{sigma-sd}
\sigma_{\chi N} \approx \frac{32}{\pi}
\frac{M^2_{\chi}m^2_N}{(M^2_{\chi}+m_N)^2} \Lambda^2 J(J+1),
\end{equation}
where
\begin{equation}
\Lambda=\frac{1}{J}\left[\langle S_p\rangle \sum_{q=u,d,s} \frac{G_{A,q}}{2} 
\Delta_q^{(p)}  +   \langle S_n\rangle \sum_{q=u,d,s} \frac{G_{A,q}}{2} 
\Delta_q^{(n)}   \right].
\end{equation}
In these expressions, $J$ is the nuclear spin, and $\langle
S_{p,n}\rangle$ are the expectation values of the spin content of
protons or neutrons in the target nucleus. The quantities $\Delta_q$
are coefficients of the matrix element of the axial current in a
nucleon, with values given by $\Delta^{(p)}_u=\Delta^{(n)}_d=0.78\pm
0.02$, $\Delta^{(p)}_d=\Delta^{(n)}_u=-0.48\pm 0.02$, and
$\Delta^{(p)}_s=\Delta^{(n)}_s=-0.15\pm 0.02$.

Inserting these values into the above equations, the WIMP-proton,
spin-dependent cross section reduces to
\begin{equation}
\sigma_{\chi p} \approx \frac{6\, m^2_p}{\pi} \left[0.78 \, G_{A,u} -0.48 \, 
G_{A,d} -0.15 \, G_{A,s} \right]^2, 
\label{cancel}
\end{equation}
which, for approximately universal couplings, yields\footnote{Notice
that in the case of universal couplings there is an approximate
cancellation of terms in Eq.\ (\ref{cancel}). Departures from the
universality of $G_{A,u}$, $G_{A,d}$ and $G_{A,s}$, however, could
lead to larger cross sections than those estimated here. Considering
the axial couplings of the $Z$ boson to fermions, for example, the
opposite signs of the couplings to up and down-type fermions leads to
an elastic scattering cross section about $10^2$ times larger than
estimated in Eq.\ (\ref{sigmaA}).}
\begin{equation}
\sigma_{\chi p} \sim  10^{-7} \, \rm{pb} \times 
\left(\frac{G_{\it A,q}}{10^{-7} \, \rm{GeV}^{-2}}\right)^2.
\label{sigmaA}
\end{equation}

Currently, the strongest constraints on spin-dependent WIMP-proton
scattering come from the COUPP \cite{coupp} and KIMS \cite{kims}
collaborations, which exclude cross sections larger than $\sigma_{\chi
p}\sim10^{-1}$ pb. This limit, however, is well beyond the range
anticipated for a thermal WIMP.

\begin{figure}[t]
\centering\leavevmode
\includegraphics[width=3.5in,angle=0]{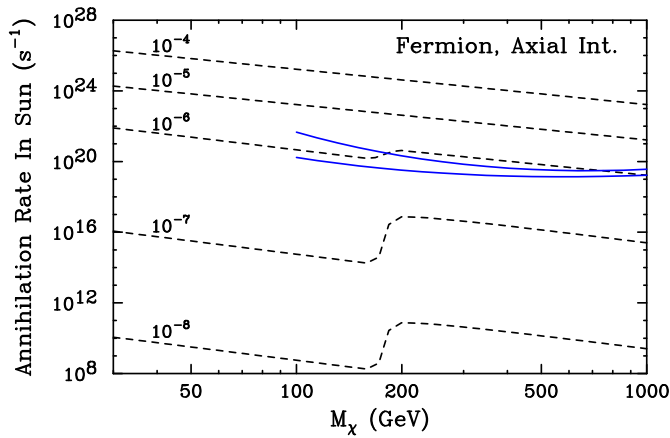}
\caption{The annihilation rate of WIMPs in the Sun, as a function of
the WIMP's mass, for a fermionic WIMP interacting with standard model
particles through axial interactions. As before, results are given for
effective couplings to each fermion species of $G_q = 10^{-8},
10^{-7}, 10^{-6}, 10^{-5}, {\rm and} \, 10^{-4}$ GeV$^{-2}$.  Also
shown as solid lines are the approximate rates needed to be
detected by an experiment such as IceCube (20 events per square
kilometer, year with a 50 GeV muon energy threshold). The solid lines
denote the reach for WIMPs annihilating to bottom quarks (top) or
gauge bosons (bottom).}
\label{fermionneutrino}
\end{figure}

In Fig.\ \ref{fermionneutrino}, we plot the annihilation rate of WIMPs
in the Sun for the case of a fermionic WIMP with axial couplings to
quarks. To be detected over the atmospheric neutrino background, the
annihilating WIMPs must generate tens of neutrino-induced muons per
year in a kilometer-scale, high-energy neutrino telescope, such as
IceCube. In Fig.\ \ref{fermionneutrino} we also plot the approximate
annihilation rate required to generate 20 events (above a muon
threshold energy of 50 GeV) per year at IceCube. This reach is shown
as solid lines for the case of WIMP annihilations to bottom quarks or
gauge bosons.

\subsection{Indirect Searches With Gamma-Rays and Charged Particles}

In addition to neutrinos, products of WIMP annihilations including
gamma rays \cite{gamma}, electrons, positrons \cite{positron,antimatter}, and
antiprotons \cite{antimatter,antiproton} could also potentially provide
detectable signals. The reach of these efforts depend on a combination
of astrophysical inputs, such as the distribution of dark matter in
the Galaxy and the properties of the galactic magnetic field, and on
the WIMP's properties, namely its mass, annihilation cross section,
and dominant annihilation modes. Although we will not, in this
article, discuss the astrophysical inputs impacting such searches, we
will briefly comment on the WIMP's annihilation cross section as it
appears in our model-independent analysis.

If we fix the WIMP's effective couplings such that its annihilation cross
section at the temperature of freeze-out is equal to the value required to yield
the observed dark matter abundance, then we can proceed to estimate its
annihilation cross section in the low-velocity limit (the relevant limit for
indirect searches). From Eqs.\ (\ref{svs}-\ref{svt}), we see that fermionic
WIMPs annihilating through pseudoscalar, vector, and tensor interactions do so largely
through terms for which $\sigma v$ is constant, rather than $\sigma v \propto
v^2$. This leads to a low-velocity annihilation cross section of approximately
$3 \times 10^{-26}$ cm$^3$/s in these cases. Scalar or axial interaction forms,
in contrast, lead to an annihilation cross section that scales as $\sigma v
\propto v^2$, and thus imply rates suppressed by a factor of about $10^{-6}$ for
WIMP annihilations in the Galactic halo.

\subsection{General Conclusions for a Fermionic WIMP}
\label{resonance}

\begin{figure}[t]
\centering\leavevmode
\includegraphics[width=3.5in,angle=0]{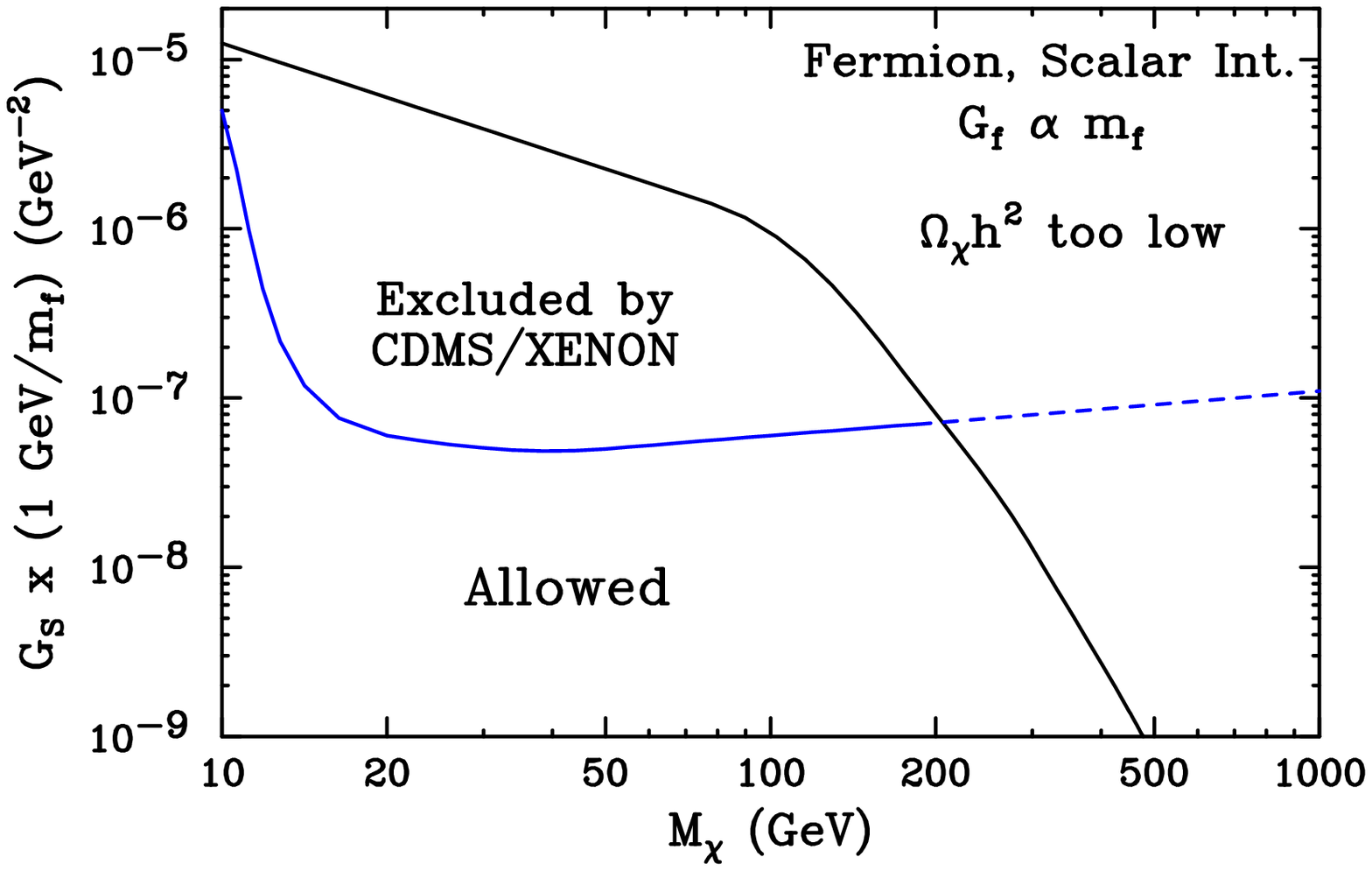}
\includegraphics[width=3.5in,angle=0]{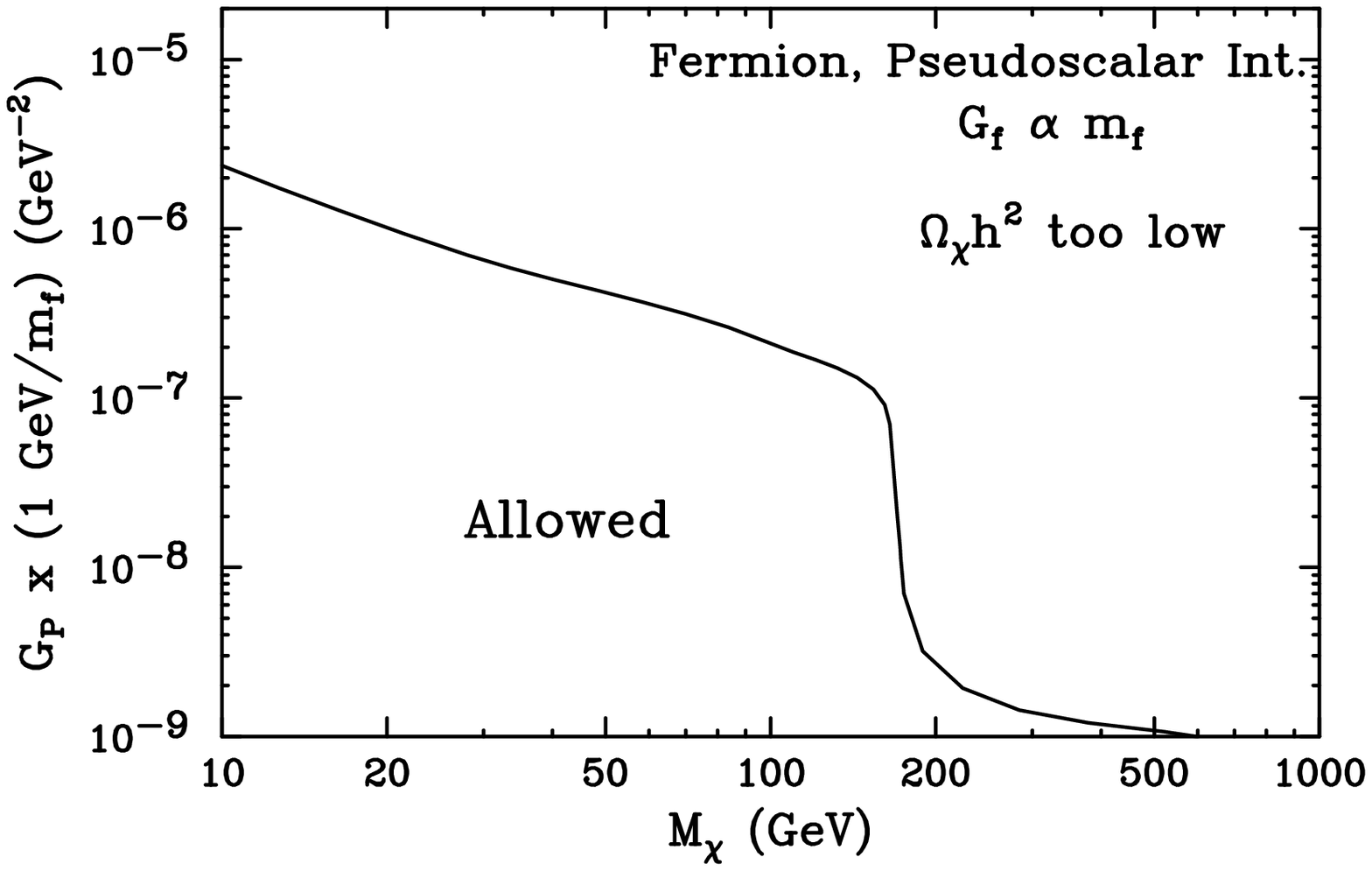}
\includegraphics[width=3.5in,angle=0]{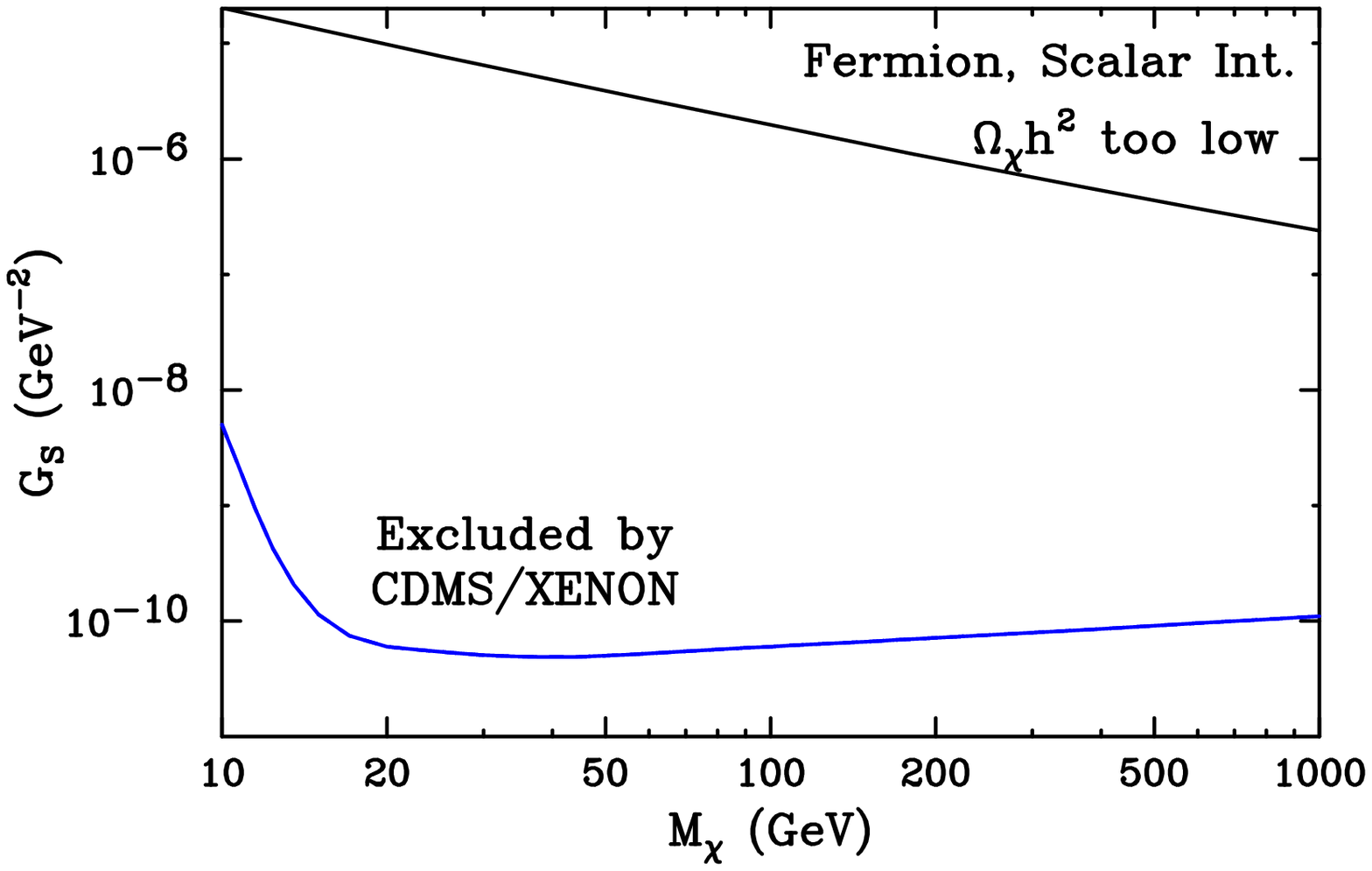}
\includegraphics[width=3.5in,angle=0]{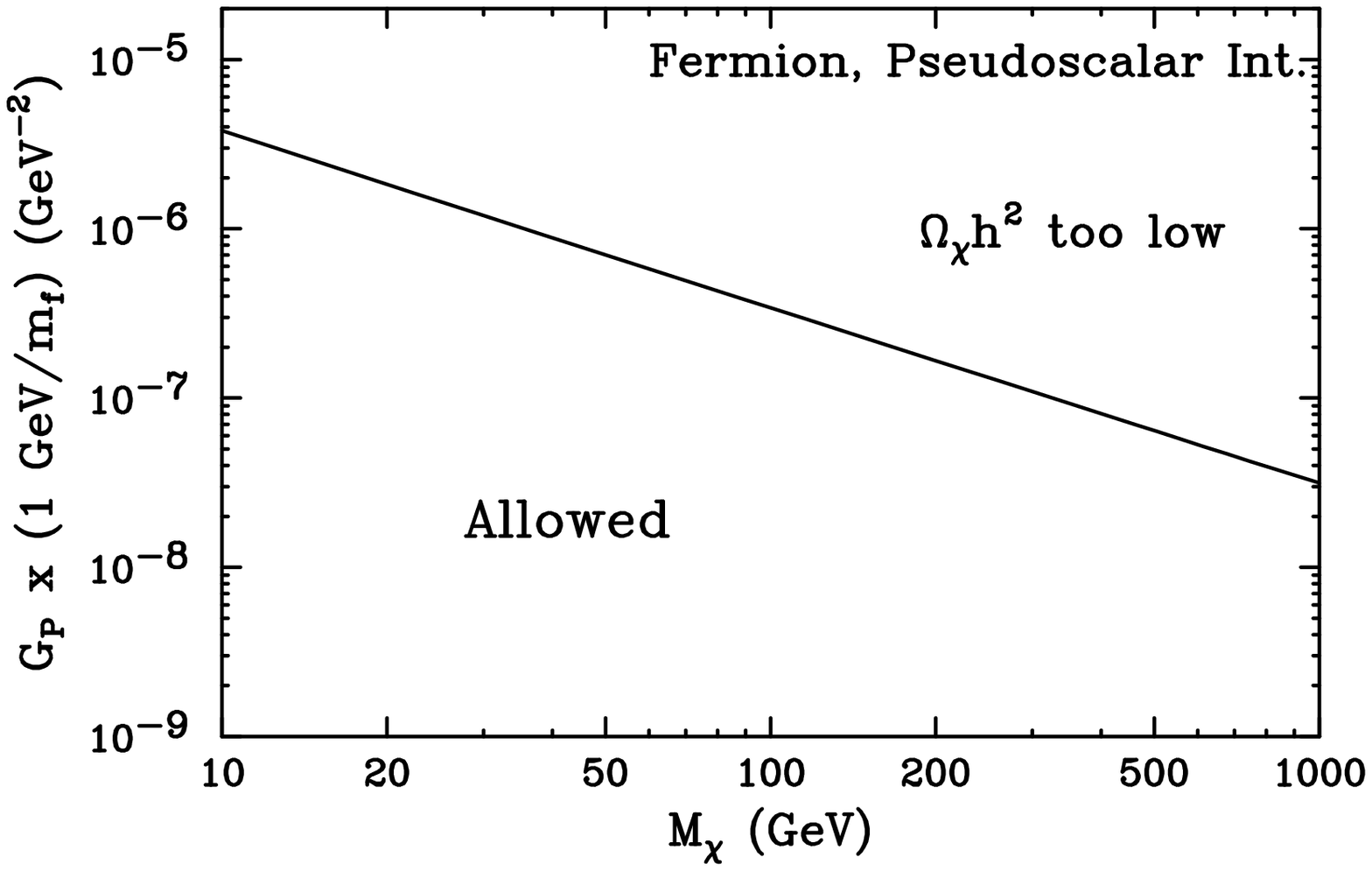}
\includegraphics[width=3.5in,angle=0]{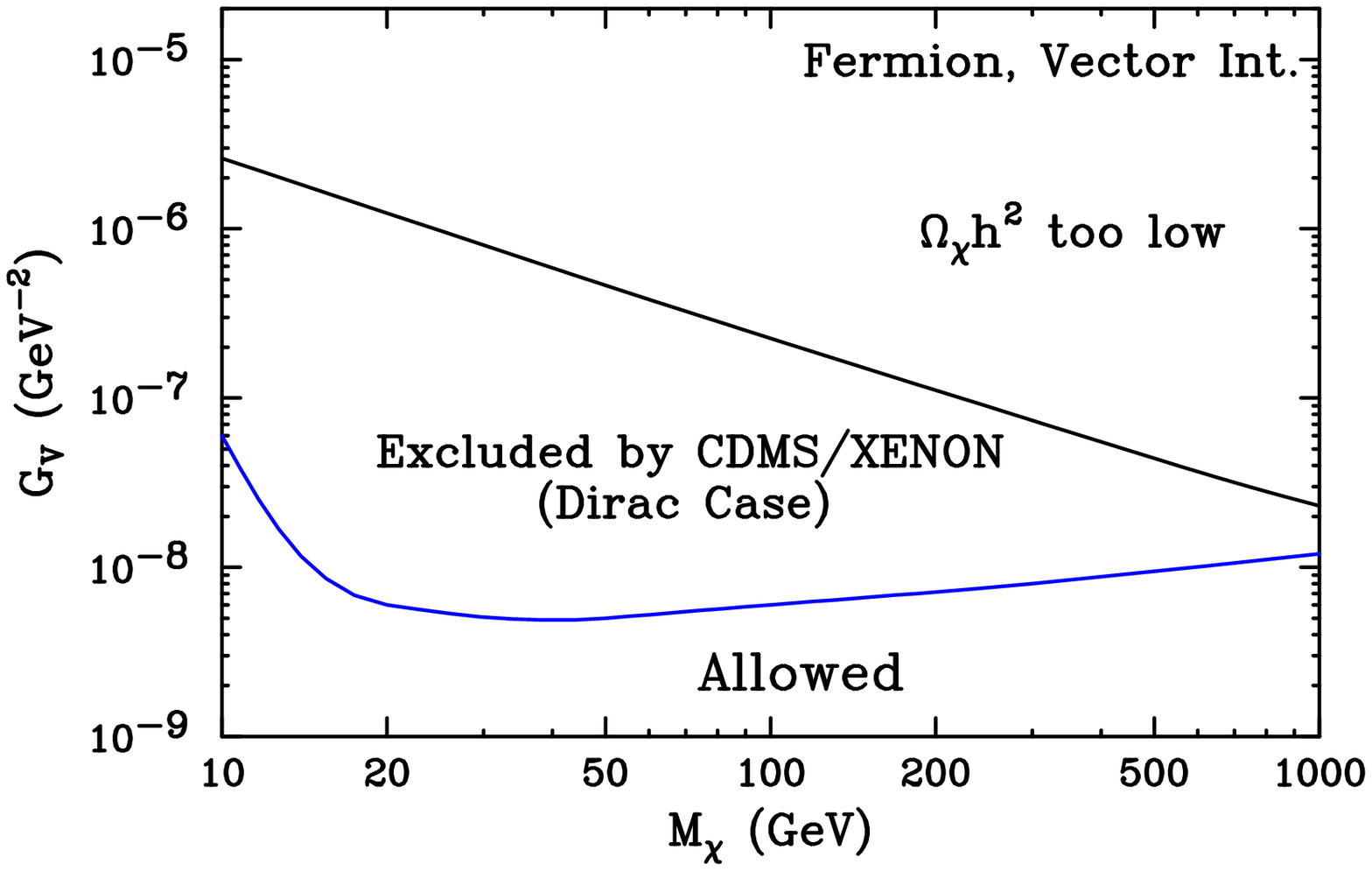}
\includegraphics[width=3.5in,angle=0]{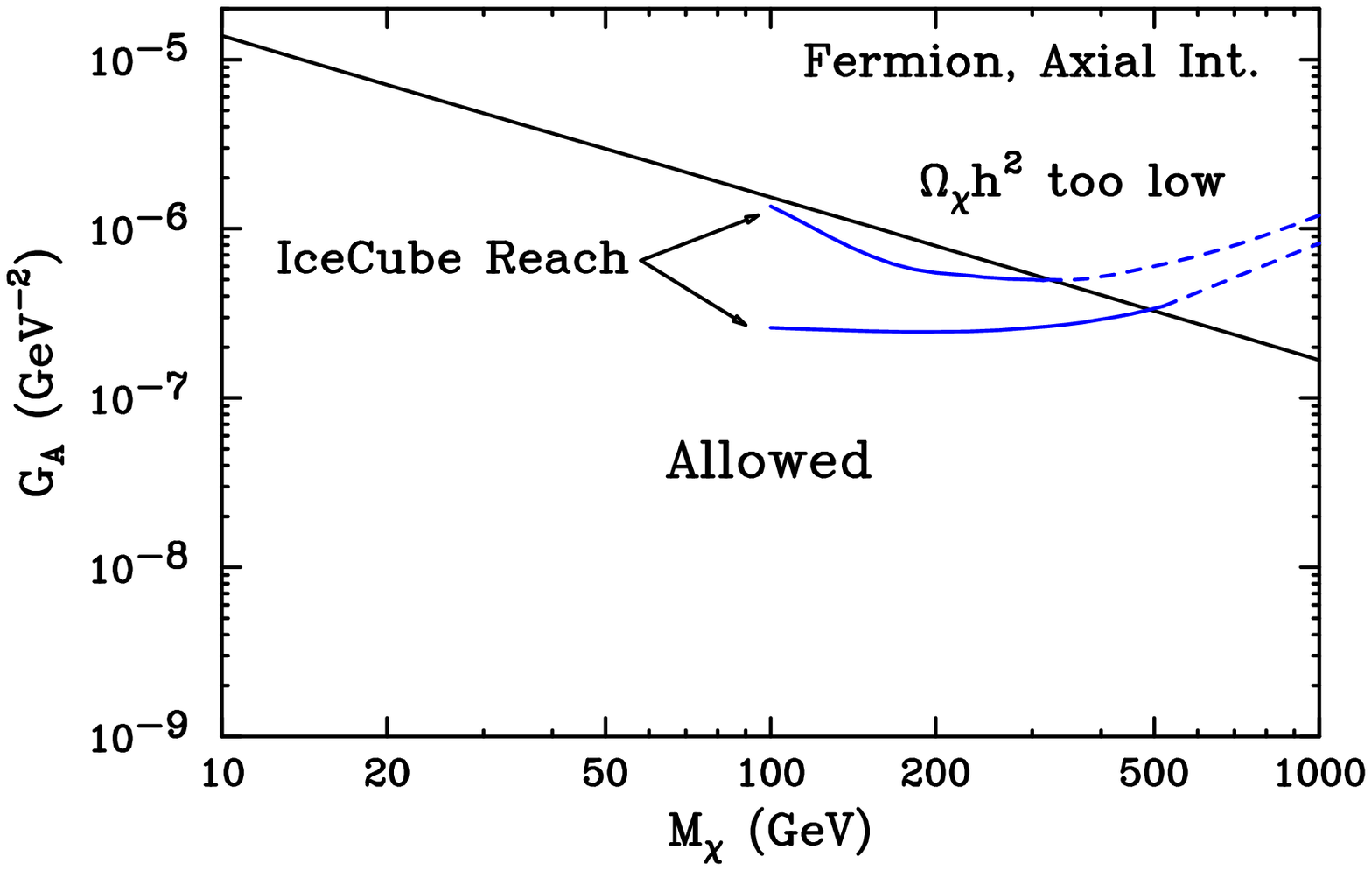}
\caption{A summary of the constraints on a fermionic WIMP with scalar,
pseudoscalar, vector, and axial interactions, including regions excluded and 
allowed by direct and indirect detection experiments (note that WIMPs with 
pseudoscalar and axial interactions are unconstrained by direct detection 
experiments). If resonances, coannihilations, or annihilations to final states 
other than fermion-antifermion pairs are significant, smaller couplings than
those shown here can lead to the measured relic abundance. See the
text for more details.}
\label{fermionsummary}
\end{figure}

Our model-independent results for a fermionic WIMP are
summarized in Fig.\ \ref{fermionsummary}. In each frame, the solid
dark (black) line denotes the combinations of WIMP masses and
couplings that lead to a thermal abundance of dark matter equal to the
value measured by WMAP \cite{wmap}. As we have pointed out,
however, these calculations were performed under the assumption that
resonances, coannihilations, and annihilations to gauge and Higgs bosons do
not play a significant role in the thermal freeze-out process. If
any of these processes have significant effects, the WIMP couplings
could be considerably smaller while still producing a dark matter
abundance consistent with WMAP.

Although Eqs.~(\ref{sigmaS}-\ref{sigmaAA}) do not form a complete set of Lorentz-invariant interaction 
Lagrangians, they are representative of a larger set of combinations of interactions:
\begin{eqnarray}
\label{lint}
\textrm{Scalar--Pseudoscalar (SP):} & \mathscr L & = \frac{G_{SP,f}}{\sqrt{2}}
\bar \chi \chi \bar f \gamma_5 f \\
\textrm{Pseudoscalar--Scalar (PS):} & \mathscr L & = \frac{G_{PS,f}}{\sqrt{2}}
\bar \chi \gamma^5 \chi \bar f f \\
\textrm{Vector--Axial Vector (VA):} & \mathscr L & = \frac{G_{VA,f}}{\sqrt{2}}
\bar \chi \gamma^{\mu} \chi \bar f \gamma_{\mu} \gamma_5 f \\
\textrm{Axial Vector--Vector (AV):} & \mathscr L & = \frac{G_{AV,f}}{\sqrt{2}}
\bar \chi \gamma^{\mu} \gamma^5 \chi \bar f \gamma_{\mu} f 
\end{eqnarray}

Up to factors of $1-m_f^2/m^2_{\chi}$, WIMPs with interactions of these types will have annihilation cross sections and relic densities determined by the WIMP interaction form (the left side of the 
Lagrangian), and elastic scattering cross sections determined by the fermion interaction form (the right side of the Lagrangian).  Consequently, interaction Lagrangians of these forms will
provide results that are redundant with what we have obtained here.  We will consider 
interaction Lagrangians of this form for scalar WIMPs, where we obtain non-redundant results. 

To roughly estimate the effect of a potential resonance, consider a WIMP
annihilating through the $s$-channel exchange of a mediator, $\psi$,
to fermion-antifermion pairs. If $m_{\psi} \gg 2 m_{\chi}$ then we can write an effective Fermi
coupling, $G_f \approx \sqrt{2} \lambda_{\chi} \lambda_f /m^2_{\psi}$,
where $\lambda_{\chi}$ and $\lambda_{f}$ are the mediator's couplings
to the WIMPs and final state fermions, respectively. If the mediator's
mass is not much greater than twice the WIMP mass, however, the
effects of the resonance on the annihilation cross section can be
significant.  In particular, we can roughly estimate an effective $G_f$ for
calculating the WIMP's annihilation cross section:
\begin{equation}
G_{f, {\rm Ann}} \approx 
\frac{\sqrt{2} \lambda_{\chi} \lambda_f}{
[(M^2_{\psi}-4 M_{\chi}^2)^2+M^2_{\psi} \Gamma^2_{\psi}]^{1/2}},
\end{equation}
where $\Gamma_{\psi}$ is the width of the mediating particle. For a
450 GeV mediator with a narrow width and a 200 GeV WIMP, the effective
value of $G_{f, {\rm Ann}}$ is a factor of about 5 larger than is
found neglecting the effects of the resonance, which enables the
measured dark matter abundance to be generated with a product of
couplings ({\it i.e.}, $\lambda_{\chi}\lambda_{f}$) that is smaller by a
factor of 5 than those shown to be required in Fig.\
\ref{fermionsummary}. In other words, the effective value of $G_f$ for
the purposes of calculating the WIMP annihilation cross section (but
not for calculating elastic scattering cross sections) is increased by
a factor of 5 in this case. For the same 450 GeV mediator and a 220 GeV
WIMP, the resonance enhances the effective value of $G_f$ by a factor
approximately 20. Although one should integrate the cross section over the thermal distribution to accurately account for the effect of a resonance on the relic abundance~\cite{Griest:1990kh}, we provide this estimate to illustrate how such a feature is qualitatively expected to impact the resulting dark matter density.

This effect is important in interpreting the constraints from direct
detection experiments and the reach of neutrino telescopes shown in
Fig.\ \ref{fermionsummary}. Consider, for example, the case of scalar
interactions with $G_f \propto m_f$ shown in the upper left
frame. Although by simply comparing the dark solid line to the lighter
(blue) solid line, the constraints from CDMS and XENON appear to rule out
a WIMP with the measured relic abundance unless it is heavier than
about $200$ GeV, this conclusion can be relaxed considerably if the
WIMP annihilates through a resonance. Similarly, if coannihilations or
annihilations to gauge/Higgs bosons play an important role in the
freeze-out process, the required effective couplings will be
considerably reduced as well.

Furthermore, departures from the universality of the WIMP's couplings
to fermions can also alter the results summarized here. WIMPs that
couple preferentially to light (heavy) quarks will be more (less)
significantly constrained by direct detection experiments and neutrino
telescopes. In an extreme case, we can imagine a WIMP that annihilates
almost entirely through couplings to gauge boson final states rather
than fermions, which in turn would lead to highly suppressed elastic
scattering cross sections.

To summarize our results for the case of a fermionic WIMP, we find:

\begin{itemize}

\item{Fermionic WIMPs with scalar interactions are required by direct
detection constraints to either 1) be heavier than about $200$ GeV, 2)
annihilate in the early universe through a resonance or
coannihilations, or 3) couple preferentially to leptons, heavy quarks,
or gauge/Higgs bosons. The case of universal couplings is very
strongly disfavored by current direct detection constraints (see the
middle-left frame of Fig.\ \ref{fermionsummary}).}

\item{The conclusions described for a fermionic WIMP with scalar
interactions also apply to the case of a Dirac fermionic WIMP with
vector interactions and fermionic WIMPs with tensor interactions. 
This is the reason why a heavy 4th generation Dirac neutrino, 
for example, is no longer an acceptable candidate for dark matter.}

\item{Fermionic WIMPs with uniquely pseudoscalar or axial interactions are not
constrained by direct detection experiments at this time.}

\item{Next generation, kilometer-scale neutrino telescopes will be
capable of constraining the case of fermionic WIMPs with axial
interactions.}

\end{itemize}

\subsection{Neutralinos as a Case Example of a Fermionic WIMP}

Departing for a moment from our model-independent analysis, we would like to
comment on our results as interpreted within the context of neutralinos, which
are attractive dark matter candidates in supersymmetric models
\cite{neutralino}. Neutralinos are Majorana fermions, and undergo scalar,
psuedoscalar, and axial interactions. Roughly speaking, neutralinos will be
overproduced in the early universe unless at least one of the following
conditions are met: 1) they are able to coannihilate efficiently with the
lightest stau or other superpartners (the coannihilation region), 2) they are
able to annihilate efficiently through the CP-odd Higgs boson resonance (the
$A$-funnel region), 3) they are a strongly mixed gaugino-higgsino, leading to
large couplings (the focus point region), or 4) much of the supersymmetric
spectrum is relatively light, making efficient annihilations possible (the bulk
region).

In the $A$-funnel region, neutralinos annihilate near resonance via
pseudoscalar interactions, but also elastically scatter through scalar
interactions associated with CP-even Higgs exchange (and squark
exchange), leading to a constraint similar in form to that shown in
the upper left frame of Fig.\ \ref{fermionsummary}, but with the solid
dark relic abundance contour reduced by at least one order of
magnitude or more. Both the $A$-funnel and bulk regions are beginning
to be significantly explored by direct detection experiments and, in
the absence of a positive detection, will be highly constrained in the
coming years.

In the focus point region, the neutralino's couplings are enhanced,
leading to scalar elastic scattering cross sections near the current
constraints from CDMS and XENON. Although the CDMS/XENON constraint
shown in Fig.\ \ref{fermionsummary} is somewhat weakened by the the
fact that neutralino annihilations in the focus point region proceed
largely to gauge boson final states, direct detection experiments will
essentially close the focus point region if no detection is made in
the next couple of years. Furthermore, focus point neutralinos have
sizable couplings to the $Z$ boson, leading to large spin-dependent
elastic scattering cross sections through axial interactions. As
mentioned before, the opposite sign of the $Z$'s couplings to up and
down type fermions leads to a much greater reach for IceCube than is
shown in the lower right frame of Fig.\ \ref{fermionsummary}. Hundreds
or thousands of events per year at IceCube are predicted throughout
much of the focus point region.

Finally, neutralinos in the stau coannihilation region are the least
constrained by direct and indirect searches, as their couplings can be
very small without leading to their overproduction in the early
universe.

\section{Scalar Dark matter}\label{scalar}

In this section, we consider the case of a scalar WIMP, with
scalar and vector, as well as combinations of scalar/pseudoscalar and vector/axial vector, interaction forms. In analogy with Eqs.\
(\ref{lint}--\ref{fermionlag}), we write
\begin{eqnarray}
\label{lint-scalar-s}
\textrm{Scalar (S):} & \mathscr L & = \frac{F_{S,f}}{\sqrt{2}}\bar \phi \phi \bar f f \\
\label{lint-scalar-v}
\textrm{Vector (V):} &\mathscr L &=\frac{F_{V,f}}{\sqrt{2}} \bar \phi \tensor{\partial}\!\!_\mu
\phi \bar f  \gamma_{\mu} f \\
\textrm{Scalar--Pseudoscalar (SP):} & \mathscr L & = \frac{F_{SP,f}}{\sqrt{2}}\bar \phi \phi \bar f \gamma_5 f \\
\textrm{Vector--Axial Vector (VA):} &\mathscr L &=\frac{F_{VA,f}}{\sqrt{2}} \bar \phi \tensor{\partial}\!\!_\mu
\phi \bar f  \gamma_{\mu} \gamma_5 f,
\end{eqnarray}
where $\phi$ denotes the scalar WIMP.  Note that in $F_f$ has mass dimension 
of $-1$ for the scalar and scalar-pseudoscalar interactions, and $-2$ for vector 
and vector-axial vector interactions.

\subsection{Scalar WIMP Annihilation and Relic Density} 

For scalar WIMPs, the annihilation cross sections to
fermion-antifermion pairs are given by
\begin{eqnarray} 
    \sigma_S  & = & \frac{1}{16\pi}
    \sum_f F_{S,f}^2 \, c_f \sqrt{\frac{s-4 m_f^2}{s-4 M_\phi^2}} \,
    \left[\frac{\, (s - 4 m_f^2)}
         {s}\right]    \\         
    \sigma_V  &=& \frac{1}{16\pi}
    \sum_f F_{V,f}^2  \, c_f \sqrt{\frac{s-4 m_f^2}{s-4 M_\phi^2}} \,
    \left[ \frac{2\, (s-4M^2_{\phi})(s+2m^2_f)}{3\,s}\right] \\
    \sigma_{SP} &=& \frac{1}{16\pi}
    \sum_f F_{SP,f}^2 \, c_f \sqrt{\frac{s-4 m_f^2}{s-4 M_\phi^2}}  \\
    \sigma_{VA} &=& \frac{1}{16\pi}
    \sum_f F_{VA,f}^2  \, c_f \sqrt{\frac{s-4 m_f^2}{s-4 M_\phi^2}} \,
    \left[ \frac{2\, (s-4M^2_{\phi})(s-4m^2_f)}{3\,s}\right] 
\end{eqnarray}
Expanding in powers of relative velocity, we arrive at
\begin{eqnarray}
    \sigma_S \vert v\vert  & \approx & \frac{1}{4\pi}\sum_f F_{S,f}^2  \, 
c_f \ \sqrt{1-  m_f^2/M^2_{\phi} } \,
    \left[\frac{1}{4}\left(1-\frac{m_f^2}{M^2_{\phi}}\right)
\right] \\
    \sigma_V \vert v\vert & \approx & \frac{1}{4 \pi}\sum_f F_{V,f}^2  \, 
c_f \  M_\phi^2 \ \sqrt{1 -  m_f^2/M^2_{\phi}}  \, 
   \left[\frac{1}{3}\left(2+\frac{m^2_f}{M_{\phi^2}}\right) \, v^2 \right] \\
  \sigma_{SP} \vert v\vert  & \approx & \frac{1}{4\pi}\sum_f F_{SP,f}^2  \, 
c_f \ \sqrt{1-  m_f^2/M^2_{\phi} } \,
    \left[\frac{1}{4}\right]\\
    \sigma_{VA} \vert v\vert & \approx & \frac{1}{4 \pi}\sum_f F_{VA,f}^2  \, 
c_f \  M_\phi^2 \ \sqrt{1 -  m_f^2/M^2_{\phi}}  \, 
   \left[\frac{1}{3}\left(2-\frac{m^2_f}{M_{\phi^2}}\right) \, v^2 \right].
\label{VAscalar}
\end{eqnarray}

\begin{figure}[t]
\centering\leavevmode
\includegraphics[width=3.5in,angle=0]{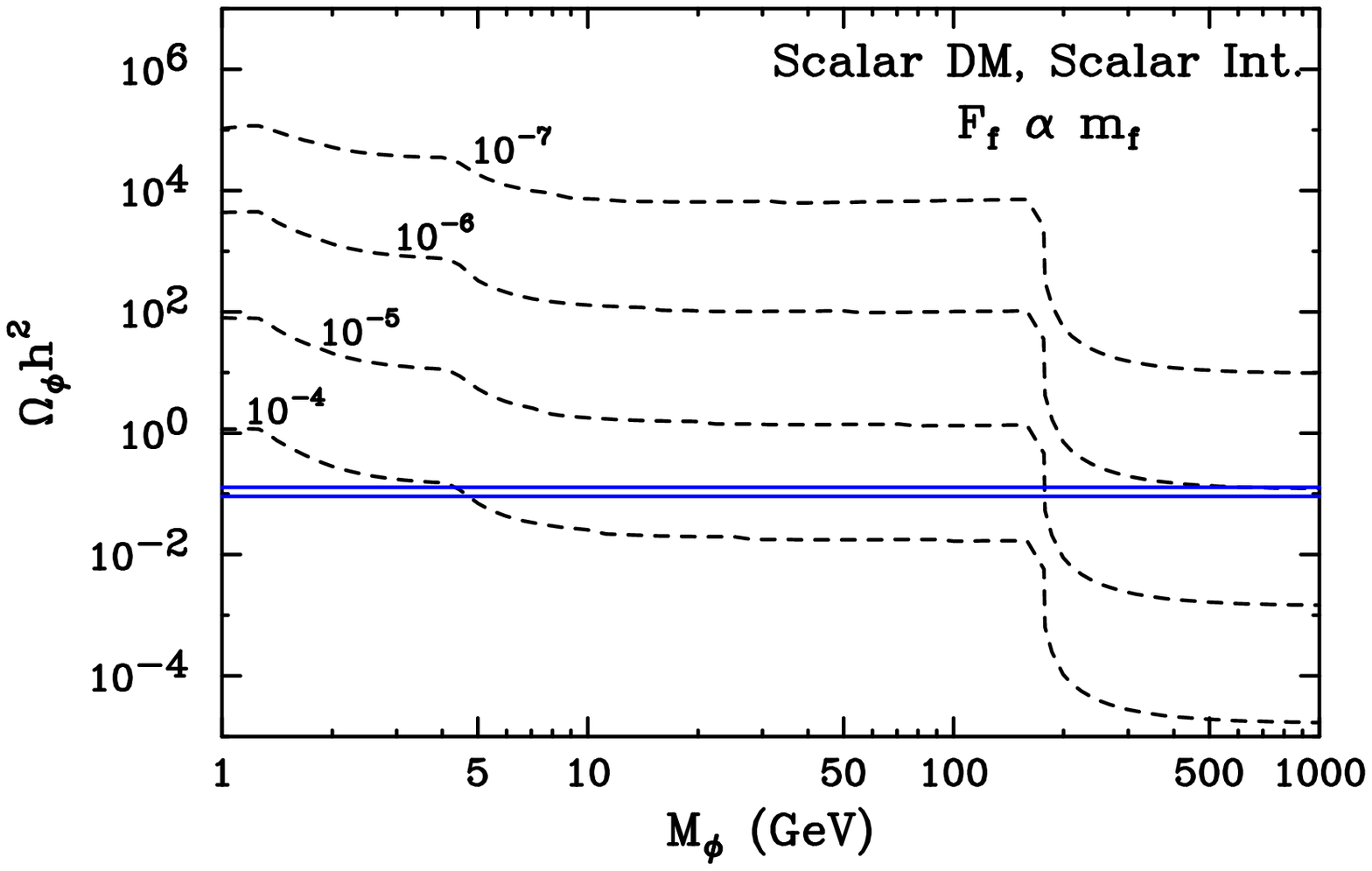}
\includegraphics[width=3.5in,angle=0]{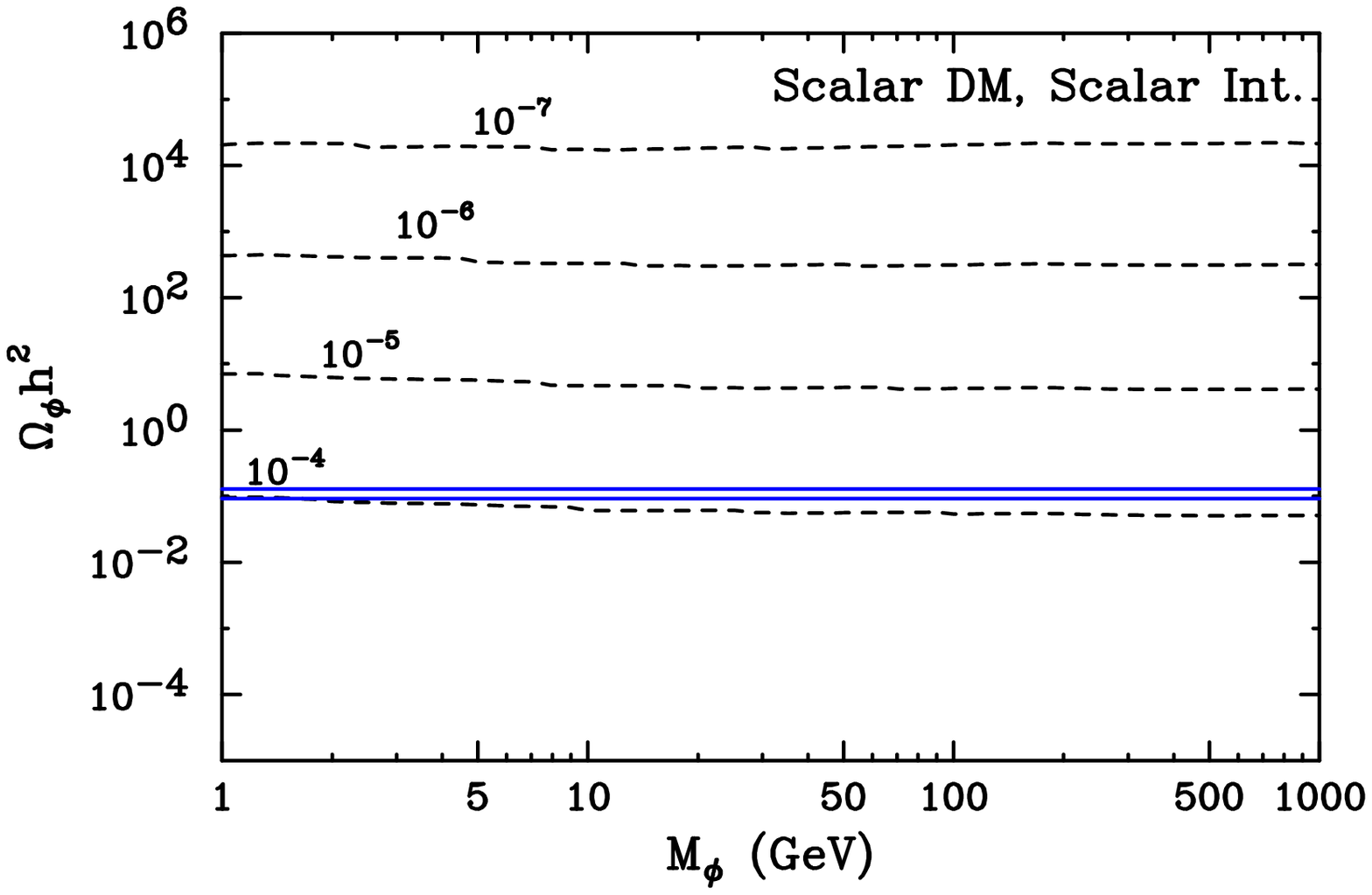}
\includegraphics[width=3.5in,angle=0]{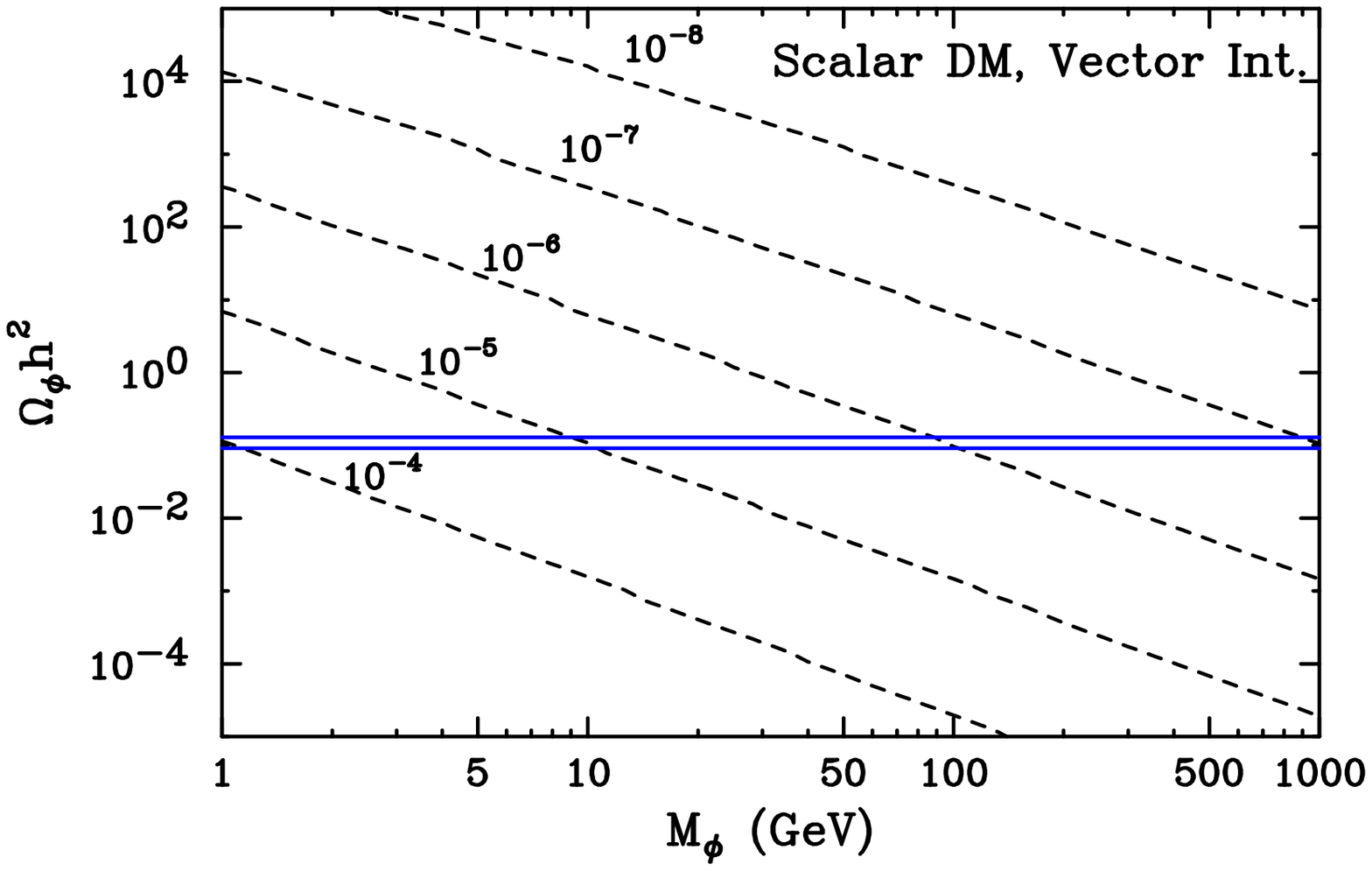}
\includegraphics[width=3.5in,angle=0]{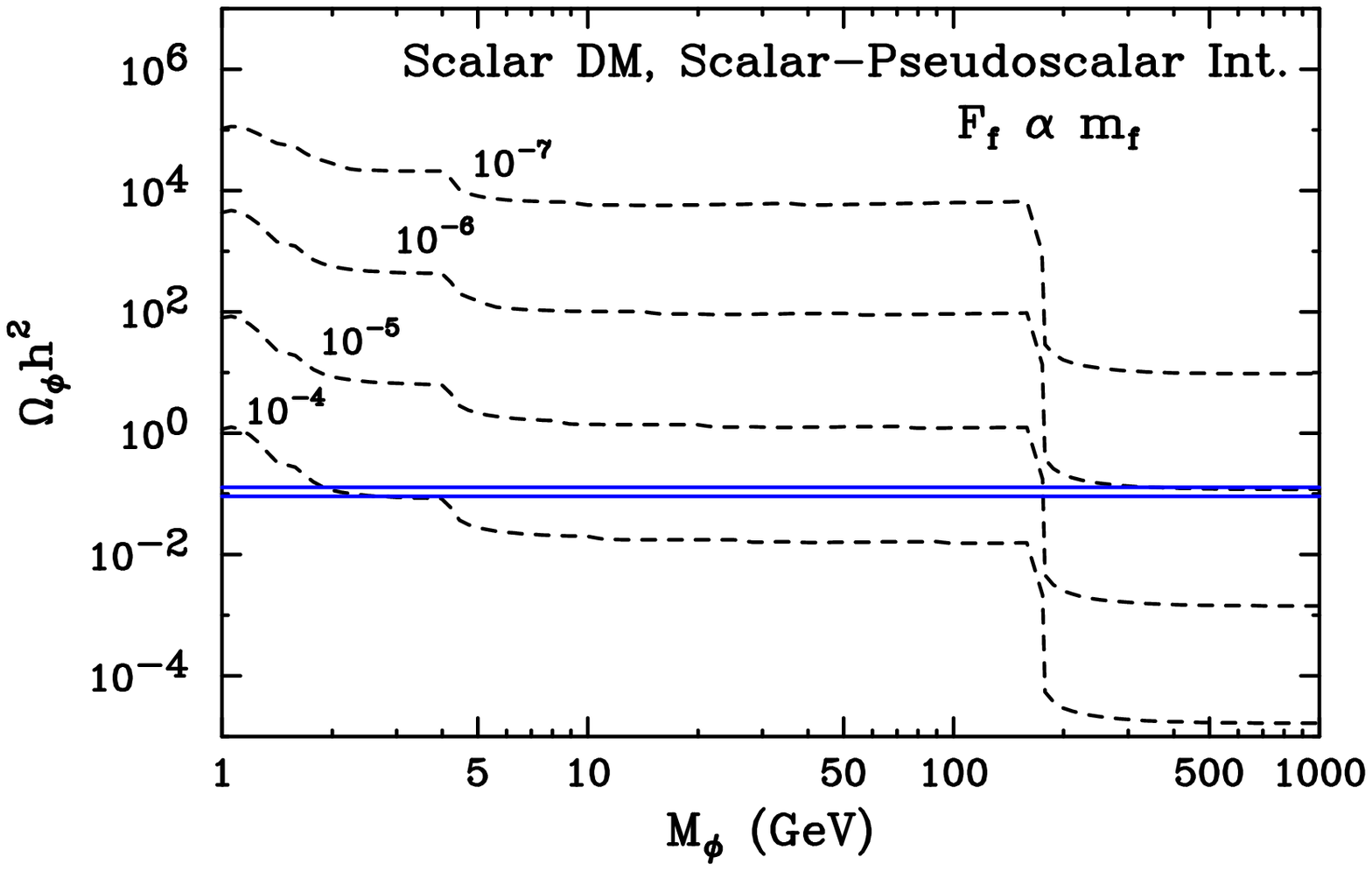}
\includegraphics[width=3.5in,angle=0]{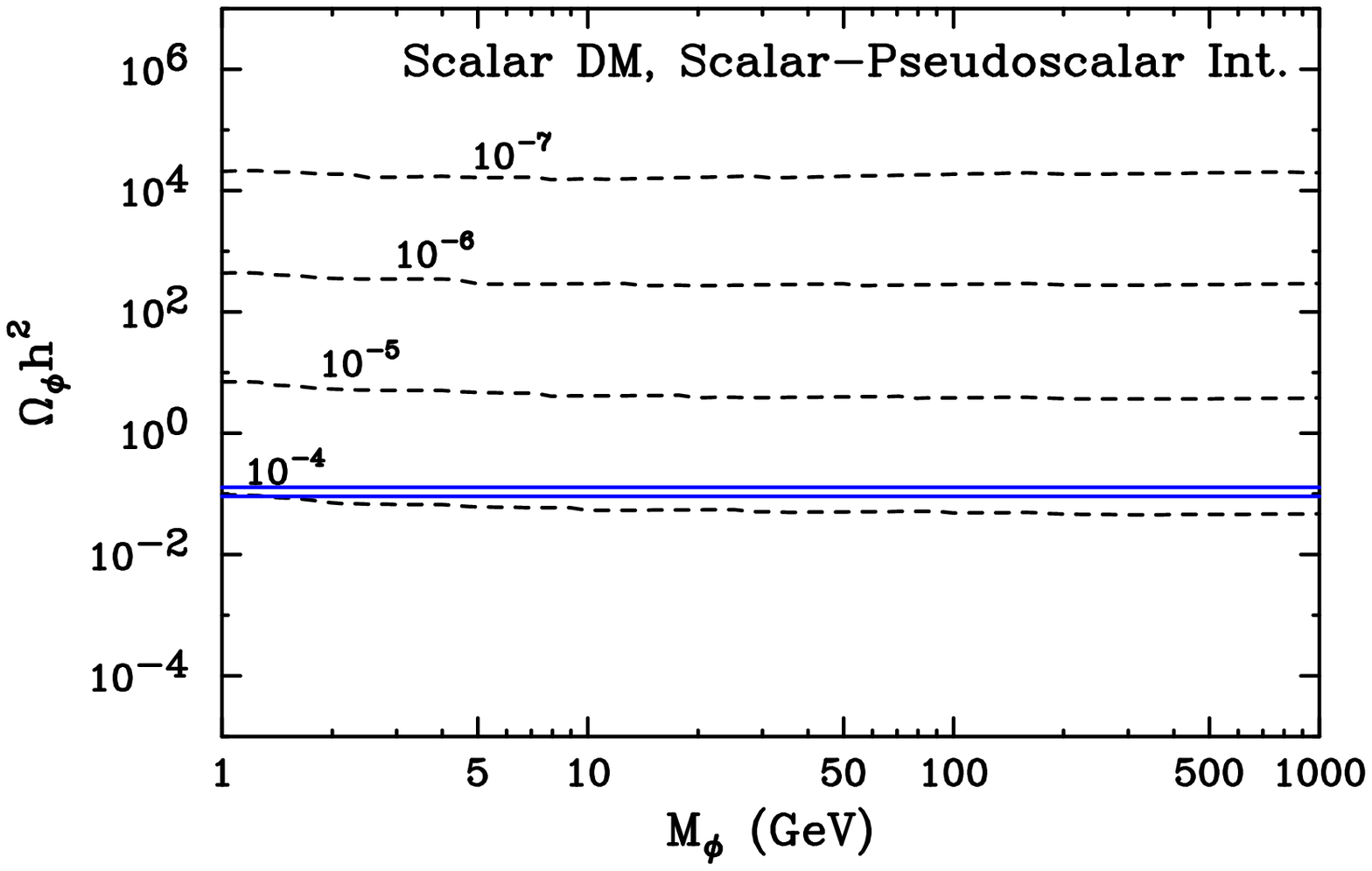}
\includegraphics[width=3.5in,angle=0]{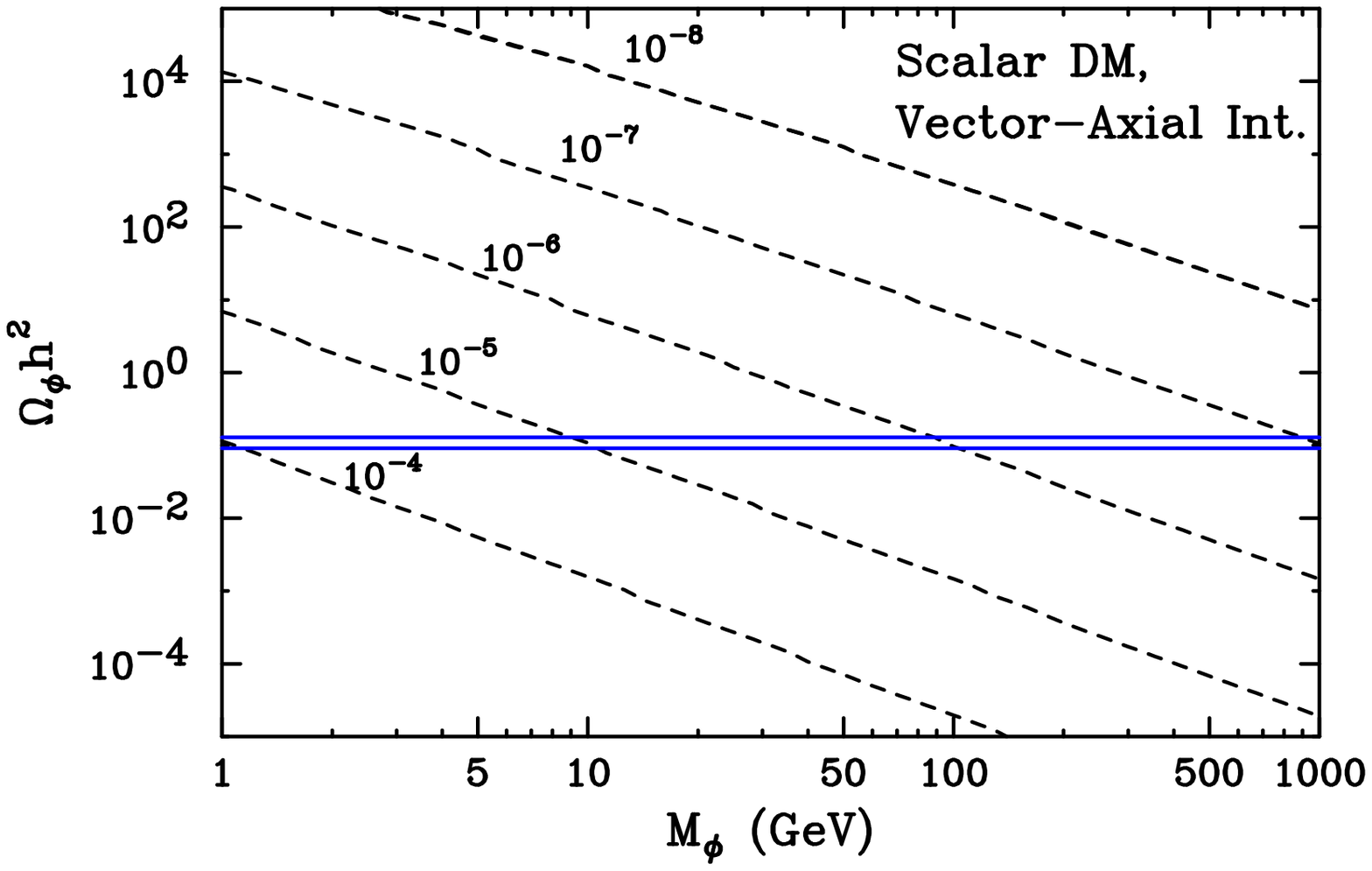}
\caption{The thermal relic density of scalar dark matter with scalar, vector, scalar-psuedoscalar, and vector-axial vector interactions with standard model particles. In the upper
left and center right frames, results are given for effective couplings to each species
of standard model fermion of $F_f \times (1\,{\rm GeV}/m_f) \times
(M_{\psi}/M_{\phi})= 10^{-7}, 10^{-6}, 10^{-5}, {\rm and} \, 10^{-4}$
GeV$^{-2}$. In the other four frames, results for $F_f = 10^{-8},
10^{-7}, 10^{-6}, 10^{-5}, {\rm and} \, 10^{-4}$ GeV$^{-2}$ are
given. If resonances, coannihilations, or annihilations to final
states other than fermion-antifermion pairs are significant, the relic
abundance is expected to be significantly lower than shown here. Also
shown as horizontal lines is the range of the cold dark matter density
measured by WMAP \cite{wmap}.}
\label{scalarrelic}
\end{figure}

To calculate the thermal relic abundance of a scalar WIMP, we follow
the same procedure as described in Sec.\ \ref{fermion}. We show the
results of this calculation in Fig.\ \ref{scalarrelic}.

\subsection{Direct and Indirect Detection}

The calculation of the elastic scattering cross section for a scalar WIMP with
nuclei is similar to that described for a fermionic WIMP in
Sec.\ \ref{directfermion}. Although we will not repeat the details of this
calculation here, we will comment on the most important differences. 

In the case of a scalar WIMP with a scalar interaction with quarks,
the effective coupling $F_q$ possesses a different dimensionality than
$G_q$. This, in turn, leads to a stronger dependence on the WIMP
mass. In particular, heavier WIMPs have a somewhat smaller elastic
scattering cross section and thus are less constrained by direct
detection experiments.

\begin{figure}[t]
\centering\leavevmode
\includegraphics[width=3.5in,angle=0]{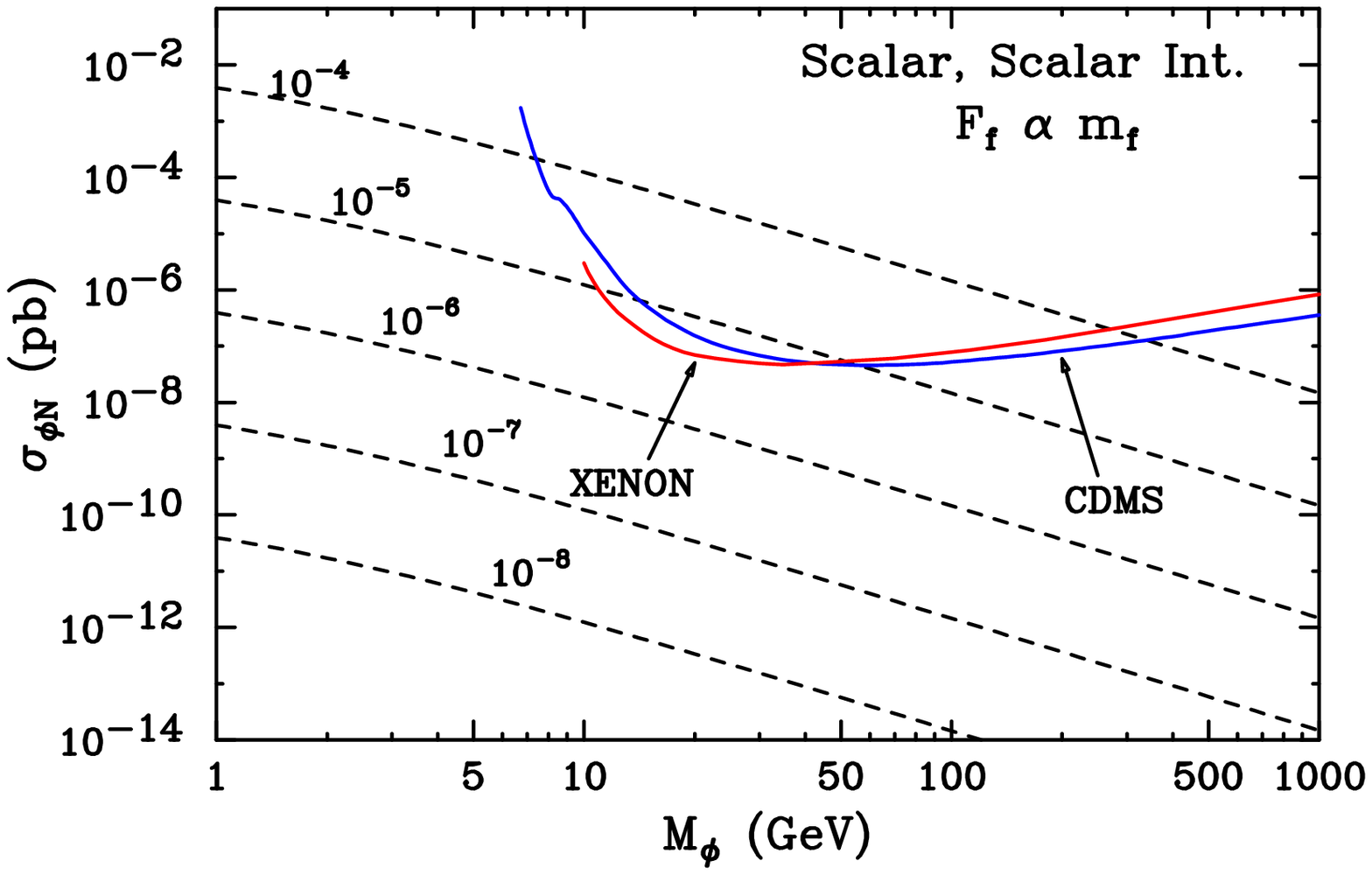}
\includegraphics[width=3.5in,angle=0]{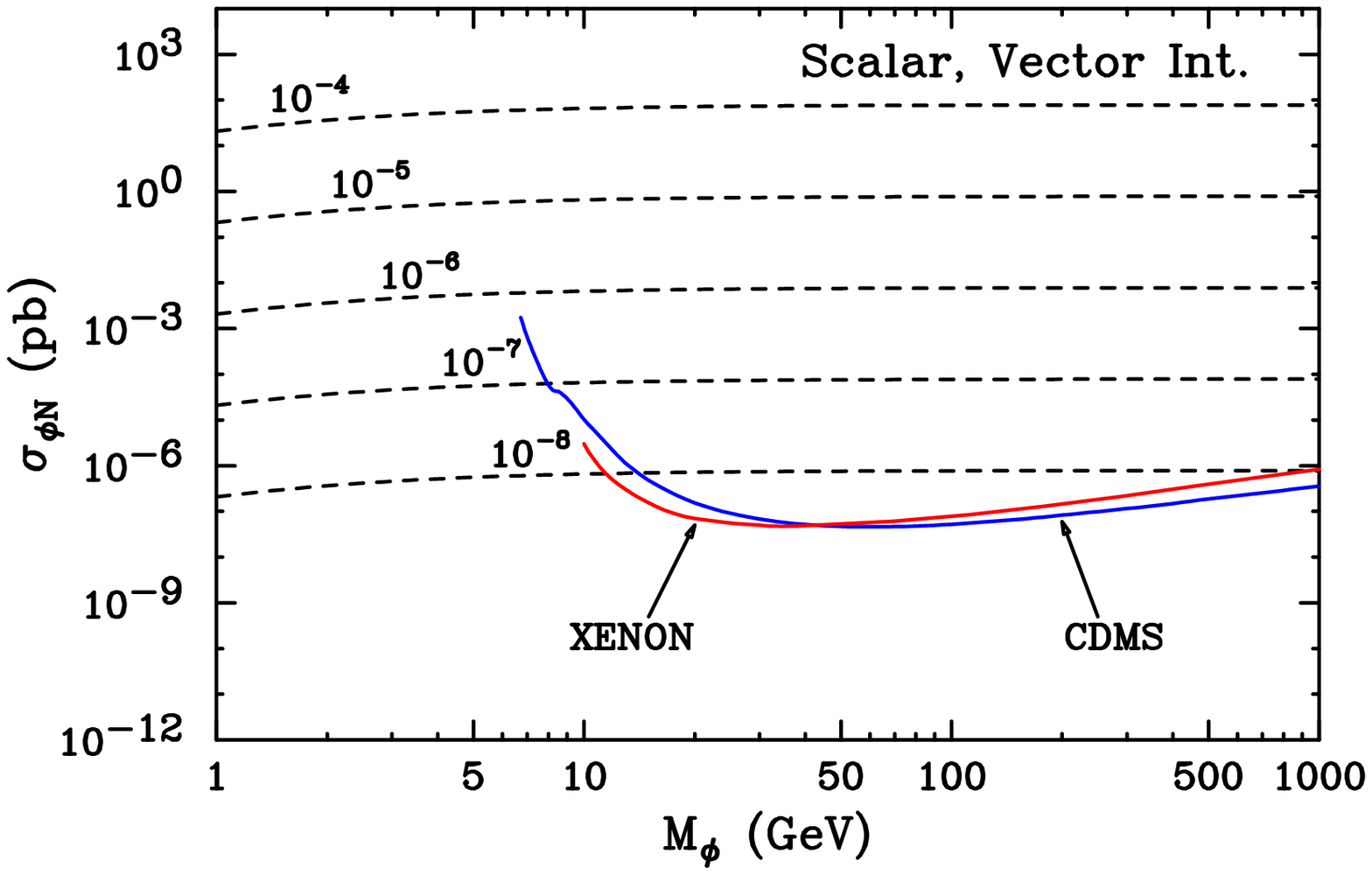}
\caption{The spin-independent WIMP-nucleon elastic scattering cross
section as a function of WIMP mass for a scalar WIMP interacting
through scalar (left) and vector (right) interactions. Results are
given for effective scalar couplings to each quark species of $F_q
\times (1\,{\rm GeV}/m_q) = 10^{-8}, 10^{-7}, 10^{-6}, 10^{-5}, {\rm
and} \, 10^{-4}$ GeV$^{-2}$ and for effective vector couplings to each
quark of $F_q = 10^{-8}, 10^{-7}, 10^{-6}, 10^{-5}, {\rm and} \,
10^{-4}$ GeV$^{-2}$. Also shown as solid curves are the current upper
limits from the CDMS \cite{cdms} and XENON \cite{xenon} experiments.
We do not show the case in which the scalar couplings are equal for
each quark species, as its leads to much larger cross sections and are
strongly excluded.}
\label{scalarSI}
\end{figure}

The elastic scattering cross sections for a scalar WIMP are shown in
Fig.\ \ref{scalarSI}. By comparing Figs.\ \ref{scalarrelic}
and \ref{scalarSI}, we can see that scalar interactions of the form
$F_f \propto m_f$ that lead to an acceptable relic density also exceed
direct detection constraints if $M_{\phi} \alt m_t$. For WIMPs heavier
than the top quark, smaller couplings allow a WIMP to evade current
direct detection constraints while also yielding an acceptable dark
matter abundance. As in the fermionic case, we find that a scalar WIMP
that annihilates largely through universal scalar couplings (equal for
all fermion species) will be essentially excluded by existing direct
detection constraints.

Scalar WIMPs with vector interactions are also severely constrained by
present direct detection experiments. By comparing the lower frames of
Figs.\ \ref{scalarrelic} and \ref{scalarSI}, we find that scalar WIMPs
with vector interactions with fermions must be heavier than several
TeV to evade current elastic scattering bounds, unless resonances,
coannihilations or annihilations to gauge/Higgs bosons play an
important role, in which case lighter WIMPs may also be allowed.

Spin-dependent scattering between scalar WIMPs and nuclei occurs only in the case of vector-axial vector interactions. Although the capture rate of scalar WIMPs in the Sun may potentially be large in this case, the annihilation cross section scales with $v^2$ (see Eq.~\ref{VAscalar}), thus suppressing the annihilation rate in the Sun's core, and along with it the resulting high-energy neutrino flux. Scalar WIMPs are, therefore, not expected to be within the reach of IceCube
or other planned high-energy neutrino telescopes.

The prospects for the indirect detection of scalar WIMPs using
gamma-rays or charged particles ($e^{\pm}$, $\bar{p}$) once again
depend on the relationship between the WIMP's annihilation cross
section and relative velocity. In the case of scalar couplings, the
annihilation cross section, $\sigma v$, is nearly independent of of
the WIMPs' relative velocity, whereas vector interactions yield
$\sigma v \propto v^2$. As a result, the indirect detection prospects
for a scalar WIMP with vector interactions are expected to be highly
suppressed.

\subsection{General Conclusions for a Scalar WIMP}

\begin{figure}[t]
\centering\leavevmode
\includegraphics[width=3.5in,angle=0]{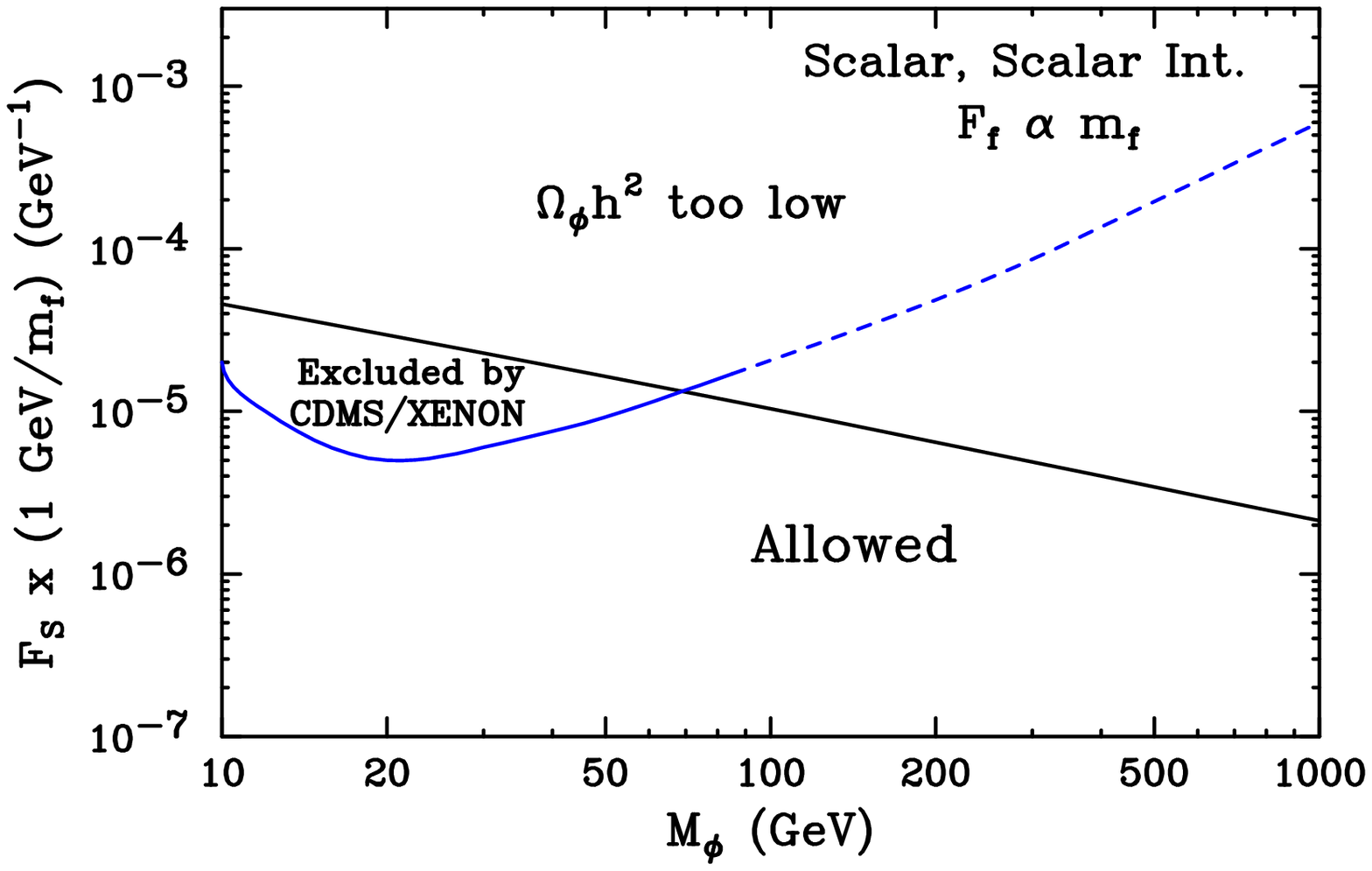}
\includegraphics[width=3.5in,angle=0]{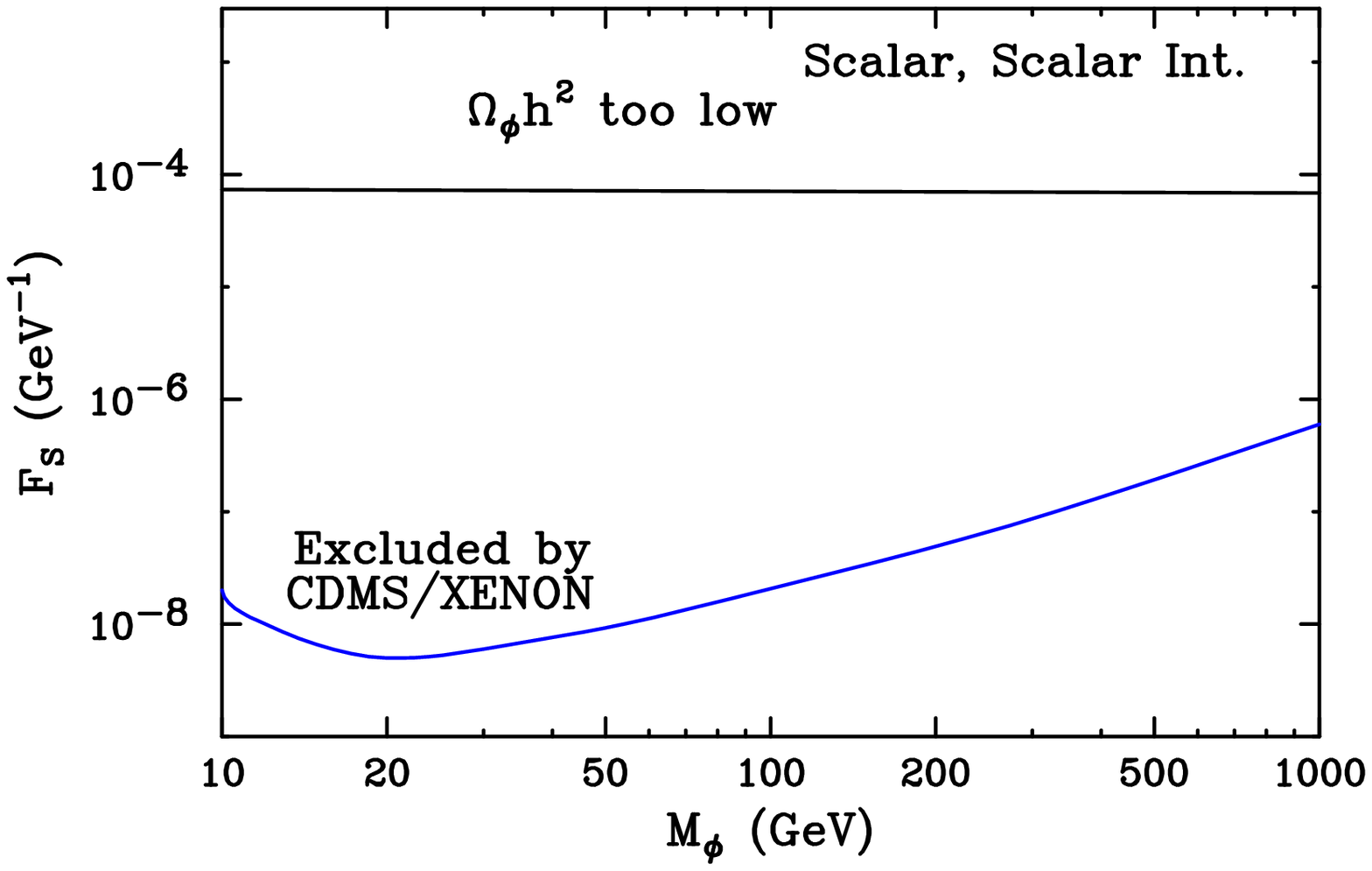}
\includegraphics[width=3.5in,angle=0]{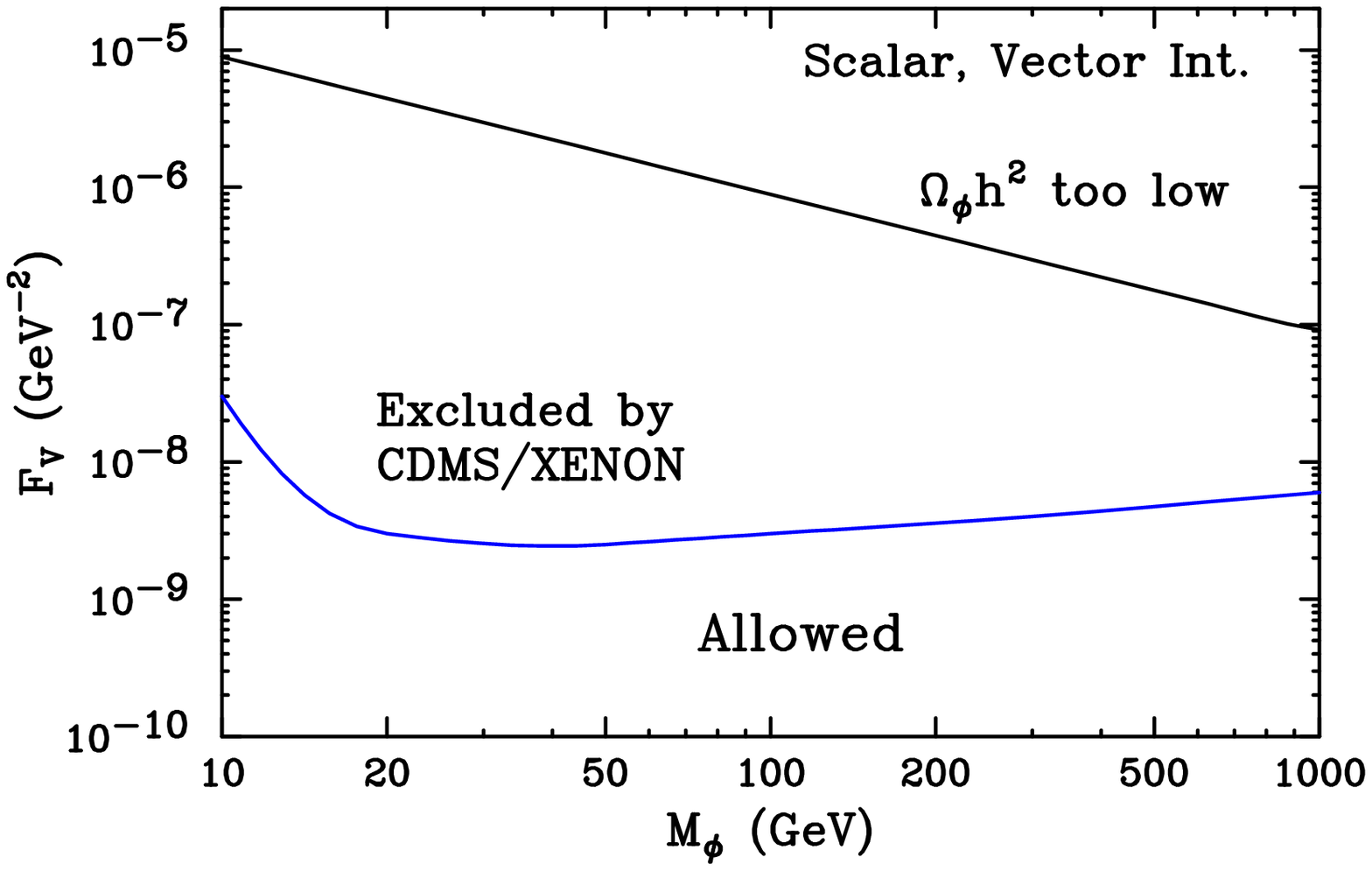}
\includegraphics[width=3.5in,angle=0]{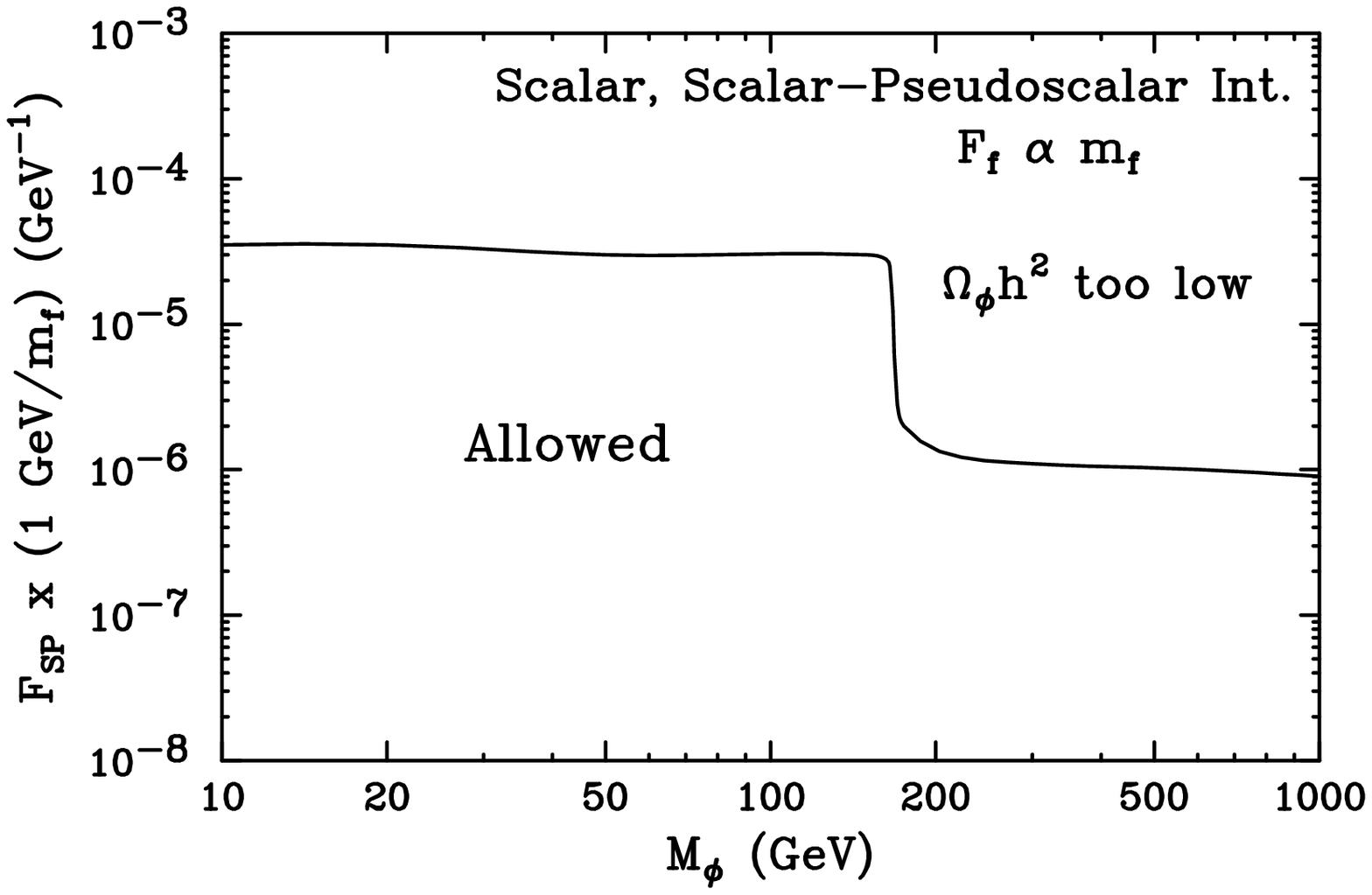}
\includegraphics[width=3.5in,angle=0]{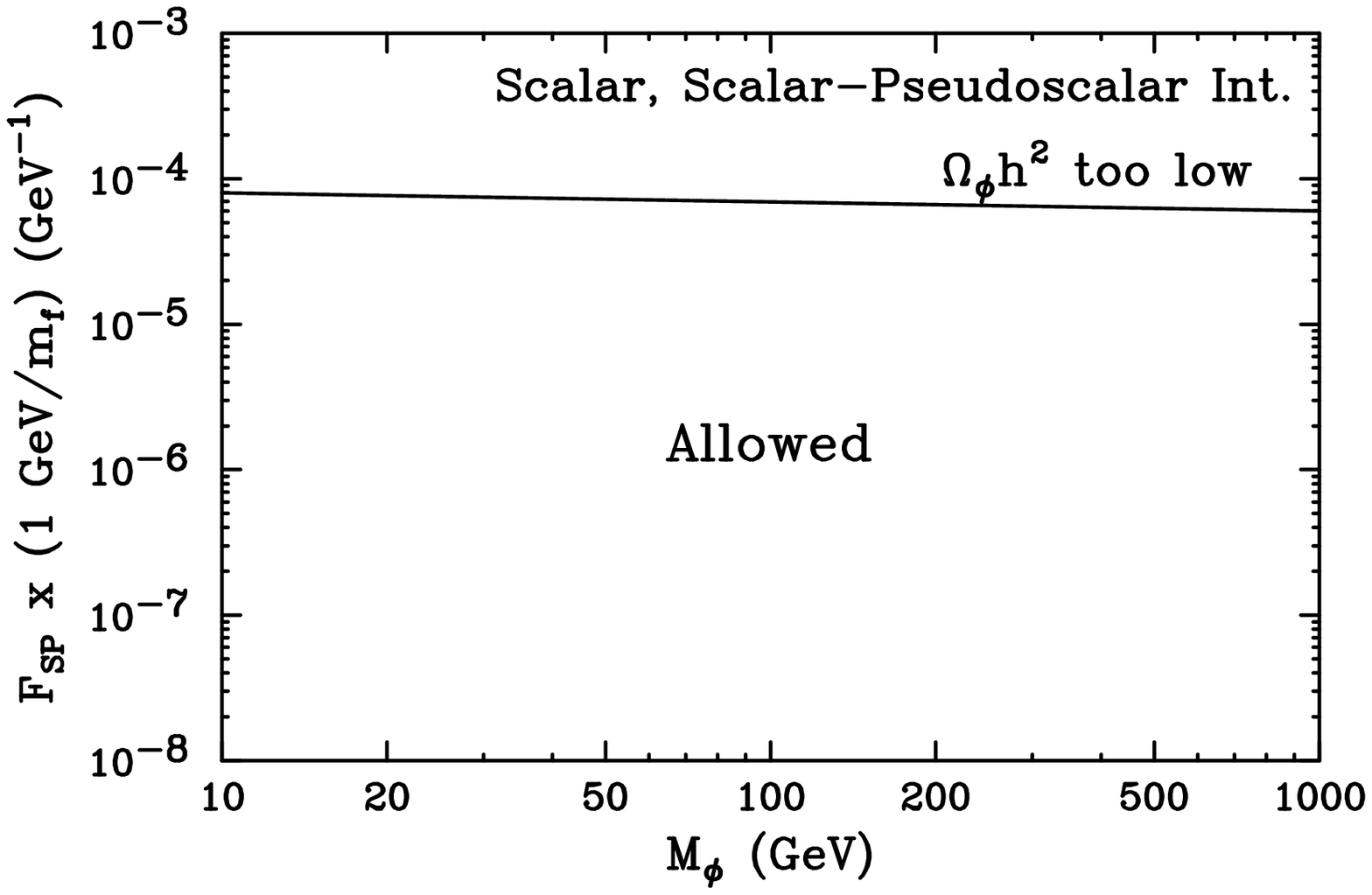}
\includegraphics[width=3.5in,angle=0]{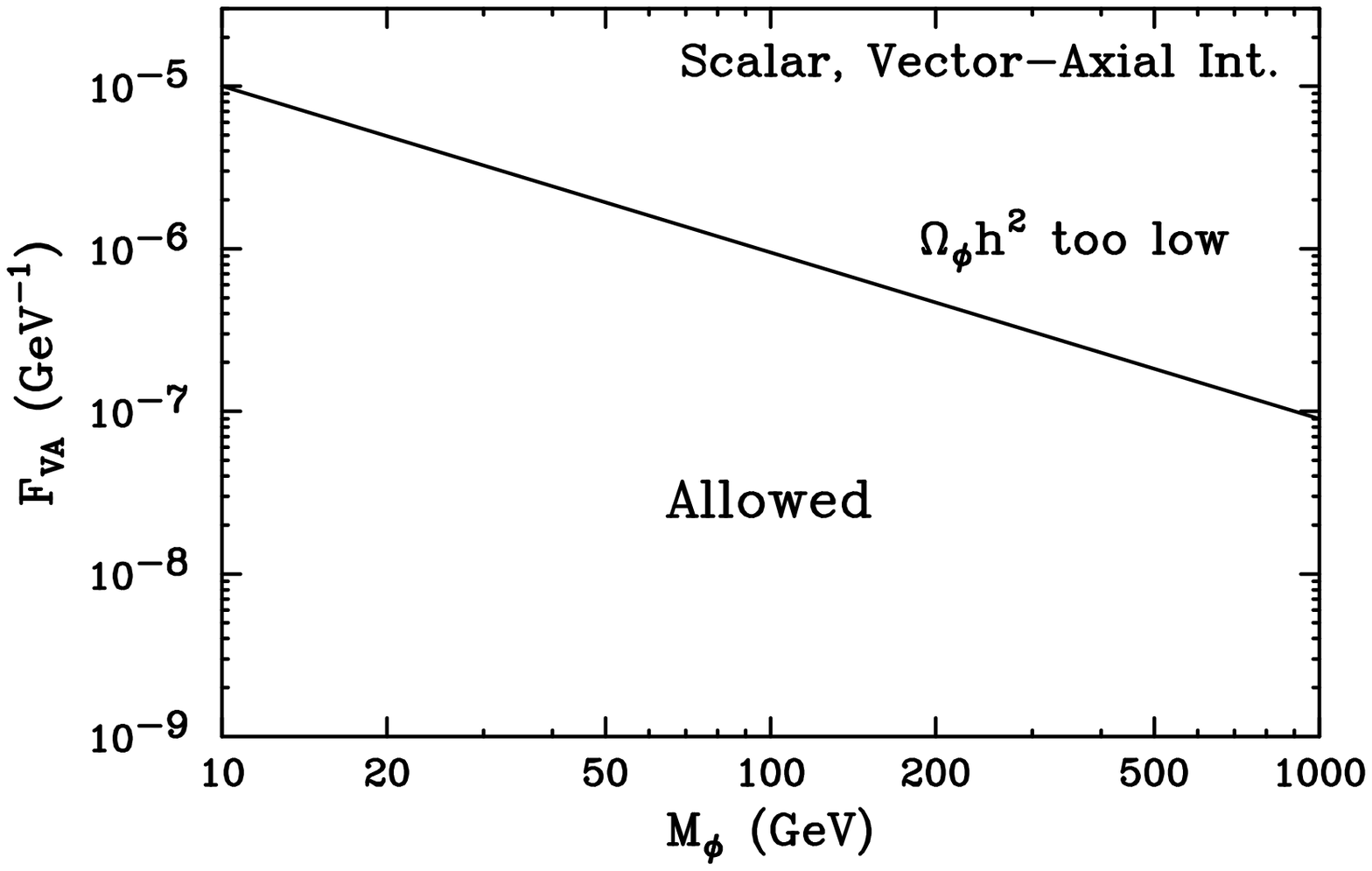}
\caption{A summary of the constraints on a scalar WIMP with scalar or
vector interactions. If resonances, coannihilations, or annihilations
to final states other than fermion-antifermion pairs are significant,
smaller couplings than those shown here can lead to the measured relic
abundance. See the text for more details.}
\label{scalarsummary}
\end{figure}

Our model-independent results for a scalar WIMP are summarized in
Fig.\ \ref{scalarsummary}. In each frame, the solid dark (black) line
denotes the combinations of WIMP mass and couplings that lead to a
thermal abundance equal to the measured dark matter density, again in
the absence of significant effects of resonances, coannihilations, or
annihilations to gauge/Higgs bosons. The lighter (blue) curve in each
frame denotes the current constraints from the direct detection
experiments CDMS and XENON.

To summarize our results for the case of a scalar WIMP, we find:

\begin{itemize}

\item{Scalar WIMPs with scalar interactions with standard model
fermions are required by direct detection constraints to 1) be heavier
than about $80$ GeV, 2) annihilate in the early universe through a
resonance or coannihilations, or 3) couple preferentially to leptons,
heavy quarks, or gauge/Higgs bosons. The case of universal couplings
is very strongly disfavored by current direct detection constraints
(see the upper-right frame of Fig.\ \ref{scalarsummary}).}

\item{The conditions described for a scalar WIMP with scalar
interactions also apply to the case of a scalar WIMP with vector
interactions. In the absence of resonances, coannihilations, and/or
annihilations to gauge/Higgs bosons, current direct detection
constraints exclude such a dark matter candidate by multiple orders of
magnitude.}

\item{Neutrino telescopes are not likely to constrain scalar WIMPs
beyond the level already achieved by direct detection experiments.}

\end{itemize}

\section{Summary and Conclusions}\label{conclusion}

Even if the Large Hadron Collider does not reveal physics beyond the
standard model, a dark matter candidate in the form of a weakly
interacting massive particle may still exist. In this article, we have
studied how the nature of such a WIMP could be deduced by its
signatures in astrophysical experiments. In our analysis, we have
taken a general and model-independent approach, considering fermionic
or scalar WIMPs with a variety of interaction forms.

In Table \ref{t1}, we summarize our findings. For each combination of
spin and interaction form, we indicate the constraints placed by and
the prospects for direct detection experiments, high-energy neutrino
telescopes, and indirect detection experiments using gamma-rays or
charged particles. Under the column of direct detection, we use the
phrases ``strongly excluded,'' ``weakly excluded,'' or ``within near
future reach,'' to denote the sensitivity or prospects for each
case. By ``strongly excluded,'' we indicate instances in which the
effective couplings to quarks, as relevant to elastic scattering with
nuclei, must be suppressed by more than a factor of ten relative to
the value required to generate thermally the observed dark matter
abundance. As we have discussed, such a suppression could result from
resonant annihilations, coannihilations, or annihilations to
gauge/Higgs bosons. The label ``weakly excluded,'' in contrast,
indicates only that the case is excluded if the effective couplings to
quarks are not suppressed by such effects. Lastly, the label ``within
near-future reach'' indicates an elastic scattering cross section
(without suppression) that is within approximately two order of
magnitudes of current direct detection limits.

Under the column of neutrino telescopes, we classify each case as
either not sensitive or sensitive over a range of WIMP masses (for
next generation experiments, such as IceCube).  This evaluation
depends on the annihilation products of the WIMP, however, and thus
are highly approximate. Under the column of gamma-rays and charged
cosmic rays, we simply indicate whether the WIMP's annihilations are
or are not suppressed by the square of the WIMP's velocity. If such
velocity suppression is present, it is highly unlikely that GLAST,
PAMELA or other planned indirect detection experiments will be capable
of detecting dark matter.

This leads us to the most obvious and important question: Will the
information provided by direct and indirect detection experiments be
able to be used to infer the particle nature of the dark matter?
Although there are certainly cases in which measurements by these
experiments will not lead to an unambiguous identification, there are
many scenarios in which a great deal could be learned. For example, if
IceCube or another high-energy neutrino telescope were to observe
neutrinos from WIMP annihilations in the Sun, we would be able to
conclude that the WIMP is likely fermionic,\footnote{More precisely, we
could conclude in this case that the dark matter particle is not a
scalar. Vector WIMPs, which we have not studied in this paper, could
also potentially generate an observable flux of high-energy
neutrinos \cite{Hooper:2002gs}.} and that it possesses an axial
interaction with light quarks. By studying the precise rate observed,
one could also potentially determine whether the WIMP's axial
interaction played a dominant or only subdominant role in the process
of thermal freeze-out. This could be combined with observations from
direct detection experiments to further constrain the possible
interactions possessed by the WIMP.

As a second possible scenario, imagine that near future direct
detection experiments observe a WIMP with a mass of a few hundred GeV
and that, shortly afterward, GLAST observes a corresponding gamma-ray
signal from WIMPs annihilating in the halo.  From Table \ref{t1}, we
can see that this leads us to only three likely possibilities: the WIMP
is either a fermion with vector interactions, a fermion with pseudoscalar-scalar interactions, or a scalar with Yukawa-like scalar interactions.

Although previous studies have shown that dark-matter experiments have the 
potential to constrain the parameters of supersymmetry \cite{susydetermine} or
even to help identify the theoretical framework from which the dark matter
arises \cite{otherdetermine}, here we have demonstrated that a far more
model-independent approach can also be fruitful. In particular, without assuming
any particular theoretical framework or model, we have shown that direct and
indirect dark matter experiments can be used to considerably constrain the spin
and interactions of the dark matter, even in the absence of any discoveries at
the LHC.

\vspace{0.5cm}
 \begin{table*}[!ht]
 \hspace{0.0cm}
\begin{center}
\begin{large}
Fermionic Dark Matter
\end{large}
\end{center}
\begin{ruledtabular}
 \begin{tabular} {c c c c c c c c} 
 Interaction &\vline& Direct Detection & \,\vline \, & Neutrino Telescopes 
&\,\vline \, &  $\gamma$-rays, $e^{\pm}$, $\bar{p}$ &  \\
 \hline \hline
Scalar  &\vline& Strongly Excluded $M_{\chi} \approx 10-100$ GeV  
& \vline & Not Sensitive  & \vline &  Suppressed by $v^2$ &  \\
 ($G_f \propto m_f$) &\vline& Weakly Excluded $M_{\chi} \approx 100-200$ GeV  
&\vline  &   & \vline &  & \\
 &\vline& Within Near Future Reach $M_{\chi} \approx 200-300$ GeV  &\vline  
&   & \vline &  & \\
\hline
Scalar  &\vline& Strongly Excluded $M_{\chi} \approx 10$ GeV$-10$ 
TeV & \vline & NA & \vline&  Suppressed by $v^2$ &  \\
($G_f$ {\rm Universal}) &\vline&  & \vline &  & \vline& &  \\
\hline
Pseudoscalar  &\vline& Not Sensitive & \vline & Not Sensitive 
& \vline&  Unsuppressed &  \\
\hline
Vector/Tensor  &\vline& Strongly Excluded $M_{\chi}\approx 10-350$ GeV 
& \vline & Not Sensitive & \vline&  Unsuppressed &  \\
&\vline& Weakly Excluded $M_{\chi} \approx 350$ GeV$-2$ TeV  
& \vline  &   & \vline &  & \\
\hline
Axial   &\vline& Not Sensitive 
& \vline & Sensitive $M_{\chi} \sim 100-500$ GeV & \vline&  Suppressed by $v^2$ &  \\
\hline
Scalar-Pseudoscalar   &\vline& Not Sensitive 
& \vline & Not Sensitive & \vline&  Suppressed by $v^2$ &  \\
\hline
Pseudoscalar-Scalar   &\vline& Weakly Excluded $M_{\chi} \approx 10-180$ GeV 
& \vline & Not Sensitive & \vline&  Unsuppressed &  \\
 ($G_f \propto m_f$) &\vline& Within Near Future Reach $M_{\chi} \approx 180-800$ GeV  
&\vline  &   & \vline &  & \\
\hline
Vector-Axial   &\vline& Not Sensitive 
& \vline & Not Sensitive & \vline&  Unsuppressed &  \\
\hline
Axial-Vector   &\vline& Strongly Excluded $M_{\chi} \approx 10$ GeV$-2$ TeV 
& \vline & Not Sensitive & \vline&  Unsuppressed &  \\
&\vline& Weakly Excluded $M_{\chi} \approx 2-10$ TeV  
& \vline  &   & \vline &  & \\

 \end{tabular}
\end{ruledtabular}
\vspace{0.5cm}
\begin{center}
\begin{large}
Scalar Dark Matter
\end{large}
\end{center}
\begin{ruledtabular}
 \begin{tabular} {c c c c c c c c} 
 Interaction &\vline& Direct Detection & \,\vline \, & Neutrino Telescopes 
 &\,\vline \, &  $\gamma$-rays, $e^{\pm}$, $\bar{p}$ &  \\
 \hline \hline
Scalar  &\vline& Weakly Excluded $M_{\phi} \approx 10-70$ GeV  
& \vline & Not Sensitive  & \vline &  Unsuppressed &  \\
 ($F_f \propto m_f$) &\vline& Within Near Future Reach $M_{\phi} 
\approx 70-200$ GeV  &\vline  &   & \vline &  & \\
\hline
Scalar  &\vline& Strongly Excluded $M_{\phi} \approx 10$ GeV$-10$ 
TeV & \vline & NA & \vline&  Unsuppressed &  \\
($F_f$ {\rm Universal}) &\vline&  & \vline &  & \vline& &  \\
\hline
Vector  &\vline& Strongly Excluded $M_{\phi}\approx 10$ GeV$-1$ TeV 
& \vline & Not Sensitive & \vline&   Suppressed by $v^2$ &  \\
 &\vline& Weakly Excluded $M_{\phi} \approx 1-5$ TeV  &\vline  &   & \vline &  
& \\
\hline
Scalar-Pseudoscalar  &\vline& Not Sensitive  
& \vline & Not Sensitive  & \vline &  Unsuppressed &  \\
\hline
Vector-Axial  &\vline& Not Sensitive  
& \vline & Not Sensitive  & \vline &  Suppressed by $v^2$ &  \\
 \end{tabular}
\end{ruledtabular}
 \caption{A summary of our results, describing the sensitivity and
 prospects for the direct and indirect detection of dark matter
 particles in the various cases we have considered. See the text for
 explanations for the labels used.}
\label{t1}
 \end{table*}

The results presented in Table \ref{t1} rely upon the set of
assumptions we have adopted. It must be noted that if dark matter consists of non-thermally
produced WIMPs, or of multiple species of particles, our conclusions
could be altered considerably. Furthermore, one might worry that the
effects of resonances, coannihilations, or annihilations to
gauge/Higgs bosons, which we have largely neglected in our analysis,
might dramatically change our conclusions. To some degree, however,
the impact of such processes are encapsulated in our definitions of
``strongly excluded'' and ``weakly excluded'', as used in
Table \ref{t1}. For example, if a WIMP annihilates largely through a
narrow resonance such that twice the mass of the WIMP lies within
approximately 5\% of the exchanged particle, then the effective
couplings relevant for elastic scattering can be reduced by a factor
of ten without the WIMP being overproduced in the early universe (see
Sec.\ \ref{resonance}). This mildly (5\% or less) fine-tuned resonance
corresponds to the ``weakly excluded'' label used in the
table. Anything labeled ``strongly excluded'' would require the masses
to be tuned even more precisely to the resonance value to remain
viable. Similarly, if a significant fraction of WIMP annihilations in
the early universe proceeded to a combination of gauge or Higgs
bosons, or occurred through coannihilations with another species of
particle, the elastic scattering cross section could be
suppressed. For the scenarios we have labeled as ``strongly excluded''
to have remained hidden from direct detection experiments, however,
about $99$\% or more of the annihilations/coannihilations of WIMPs in
the early universe must have occurred through such processes. So
although the conclusions we have reached here are not entirely immune
to the inclusion of such effects, they are quite robust in all but the
most extreme cases.

In conclusion, we find that in the case that the Large Hadron Collider
does not discover physics beyond the standard model, astrophysical
experiments may still be able to constrain the nature of the dark
matter, even without assuming supersymmetry or any other specific
particle physics framework. In particular, the spin and interaction
forms of dark matter can potentially be identified by combining
results from direct detection experiments, neutrino telescopes, and
indirect detection experiments using gamma-rays or charged cosmic
rays.

\vspace{0.5cm}

\acknowledgments{We would like to thank Rakhi Mahbubani, Keith Olive, and Pearl
Sandick for illuminating discussions. DH is supported by the Fermi Research
Alliance, LLC under Contract No.\ DE-AC02-07CH11359 with the US Department of
Energy and by NASA grant NNX08AH34G.}



\end{document}